\documentclass{article}

\usepackage{arxiv}
\usepackage{orcidlink}

\title{FairStream:\\ Fair Multimedia Streaming Benchmark for Reinforcement Learning Agents}
\renewcommand{\shorttitle}{FairStream: Fair Multimedia Streaming Benchmark for Reinforcement Learning Agents}

\renewcommand{\headeright}{}
\renewcommand{\undertitle}{}
\date{}

\author{
  Jannis Weil\,\orcidlink{0000-0001-5439-9131}$^{1,*}$\\
  \texttt{jannis.weil@tu-darmstadt.de} \\
  \And
  Jonas Ringsdorf\,\orcidlink{0009-0002-6042-9960}$^{1,*}$\\
  \texttt{jonas.ringsdorf@stud.tu-darmstadt.de} \\
  \And
  Julian Barthel\,\orcidlink{0009-0006-1285-6916}$^{1}$\\
  \texttt{j.barthel96@gmx.de} \\
  \And
  Yi-Ping Phoebe Chen\,\orcidlink{0000-0002-4122-3767}$^2$\\
  \texttt{phoebe.chen@latrobe.edu.au}\\
  \And
  Tobias Meuser\,\orcidlink{0000-0002-2008-5932}$^1$\\
  \texttt{tobias.meuser@tu-darmstadt.de}\\
  \AND
  \textnormal{$^{1}$Multimedia Communications Lab (KOM), Technical University of Darmstadt, Germany} \\
  \textnormal{$^{2}$Department of Computer Science and Information Technology, La Trobe University, Australia} \\
  \textnormal{$^{*}$Both authors contributed equally to this research.}
  \vspace{-1.0em}
}

\usepackage{amssymb}
\usepackage{amsmath}

\usepackage[numbers]{natbib}
\usepackage{xspace}
\usepackage{mathtools}
\usepackage{amsmath}
\usepackage{booktabs}
\usepackage{multirow}
\usepackage{stfloats} 
\usepackage{enumitem} 
\usepackage[nolist]{acronym}

\newcommand{\rllib}{RLlib\xspace}
\newcommand{\gminerva}{Greedy-$8$-Minerva\xspace}
\newcommand{\greedy}{Greedy-$8$\xspace}

\DeclareSymbolFont{bbold}{U}{bbold}{m}{n}
\DeclareSymbolFontAlphabet{\mathbbold}{bbold}

\begin{acronym}[ECU]
\acro{qoe}[QoE]{Quality of Experience}
\acro{hmd}[HMD]{Head-Mounted Display}
\acro{rl}[RL]{Reinforcement Learning}
\acro{marl}[MARL]{Multi-Agent Reinforcement Learning}
\acro{ml}[ML]{Machine Learning}
\acro{vr}[VR]{Virtual Reality}
\acro{tcp}[TCP]{Transmission Control Protocol}
\acro{ssim}[SSIM]{Structural Similarity Index Measure}
\acro{dash}[DASH]{Dynamic Adaptive Streaming over HTTP}
\acro{abr}[ABR]{Adaptive Bitrate}
\acro{pcv}[PCV]{Point Cloud Video}
\acro{cv}[CV]{Coefficient of Variation}
\acro{ppo}[PPO]{Proximal Policy Optimization}
\acro{lstm}[LSTM]{Long Short-Term Memory}
\acro{vmaf}[VMAF]{Video Multi-Method Assessment Fusion}
\acro{acr}[ACR]{Absolute Category Rating}
\acro{a3c}[A3C]{Asynchronous Advantage Actor-Critic}
\acro{dqn}[DQN]{Deep Q-Networks}
\end{acronym}

\DeclareFontFamily{U}{FdSymbolA}{}
\DeclareFontShape{U}{FdSymbolA}{m}{n}{
    <-> s * [1] FdSymbolA-Book
}{}
\DeclareSymbolFont{fdsymbols}{U}{FdSymbolA}{m}{n}
\DeclareMathSymbol{\medblacktriangleup}{\mathbin}{fdsymbols}{83}
\DeclareMathSymbol{\medblackcircle}{\mathbin}{fdsymbols}{99}
\DeclareMathSymbol{\medblacksquare}{\mathbin}{fdsymbols}{118}
\DeclareMathSymbol{\medblackdiamond}{\mathbin}{fdsymbols}{132}
\DeclareMathSymbol{\medblackstar}{\mathbin}{fdsymbols}{149}

\usetikzlibrary{calc}
\usetikzlibrary{positioning}

\begin{document}

\maketitle

\begin{abstract}
  Multimedia streaming accounts for the majority of traffic in today's internet.
Mechanisms like adaptive bitrate streaming control the bitrate of a stream based on the estimated bandwidth, ideally resulting in smooth playback and a good \ac{qoe}.
However, selecting the optimal bitrate is challenging under volatile network conditions.
This motivated researchers to train \ac{rl} agents for multimedia streaming.
The considered training environments are often simplified, leading to promising results with limited applicability.
Additionally, the \ac{qoe} fairness across multiple streams is seldom considered by recent \ac{rl} approaches.
With this work, we propose a novel multi-agent environment that comprises multiple challenges of fair multimedia streaming: partial observability, multiple objectives, agent heterogeneity and asynchronicity.
We provide and analyze baseline approaches across five different traffic classes to gain detailed insights into the behavior of the considered agents, and show that the commonly used \ac{ppo} algorithm is outperformed by a simple greedy heuristic.
Future work includes the adaptation of multi-agent \ac{rl} algorithms and further expansions of the environment.
\acresetall
\end{abstract}

\keywords{Multi-agent Environments, Fair Multimedia Streaming, Heterogeneous Clients, Asynchronous Agents, Partial Observability, Multi-Objective Optimization}

\section{Introduction}
\ac{marl} has been widely applied in the field of communication systems, as many control problems from this field can be framed as finding an optimal policy in a multi-agent system~\cite{li2022MARLinFutureInternet}.
These problems cover a variety of challenges from basic \ac{rl} research.
We will briefly describe selected challenges in the following.

\textit{Partial observability} and \textit{multiple objectives} are very common in communication systems and frequently considered by related works in that area. 
\textit{Partial observability} is given when agents do not observe the full system state~\cite{jaakkola94rlpomdp} and can be caused by a high overhead of network monitoring, the consideration of privacy, or a lack of trust between systems.
An agent has \textit{multiple objectives}~\cite{hayes2022practical} if its overall goal consists of multiple, potentially conflicting, subgoals. 
An example would be maximizing throughput while minimizing energy consumption. 
Related works in the area of communication networks~\cite{luong19DRLApplications} often combine multiple subgoals into a single reward function,  
but seldom explore the solution space with respect to the individual objectives.

Although related works consider increasingly complex and realistic system models, some fundamental challenges of real communication systems are still largely unexplored.
In particular, there is little related work that considers \textit{agent heterogeneity} and \textit{asynchronicity}.
Agents are \textit{heterogeneous} when they have different observation spaces, action spaces, reward functions, or roles~\cite{zhong2024heterogeneous}.
This is the case whenever individual systems have different resource requirements, functionality, or goals~\cite{su24qoefairnessHeterogeneousCongestionControl, georgopoulos13openFlowQoEFair}.
\textit{Asynchronicity} is also very common in communication systems. 
Decisions are usually made asynchronously in an event-based manner, e.g., when a system receives a packet and has to make a routing decision.
In contrast, existing \ac{marl} environments and algorithms usually assume synchronous agents~\cite{yu23asyncPPO}.
Consequently, approaches for communication systems often leverage single-agent \ac{rl} methods which do not capture potential interactions between agents.

Existing \ac{marl} environments for communication systems usually address a subset of these challenges.
While the resulting assumptions allow to develop proficient solutions, they can hinder the transfer of the learned behavior to real communication systems.

In this paper, we aim to bridge this gap in the context of multimedia streaming and
propose a cooperative multi-agent environment that comprises all of the aforementioned challenges.
Each agent represents a streaming client that aims to maximize its own quality while considering fairness between all clients.
Based on the well-established \ac{dash} model~\cite{dashStandard2022}, clients download multimedia content that is split into segments of short duration, store them in a buffer and then continuously play back the stored segments. 
At each step, agents receive a \textit{partial observation} that represents the network's state and \textit{asynchronously} select the quality level of the following segment.
Agents have \textit{multiple objectives}, as the perceived streaming quality and the fairness in terms of quality differences between individual devices are jointly optimized.
In our experiments, we consider four \textit{heterogeneous} client types with different resource demands.

Our contributions are as follows:
\begin{itemize}
    \item Design of an environment for fair multimedia streaming, capturing the challenges associated with partial observability, agent asynchronicity, heterogeniety, and multiple objectives.
    \item Provisioning of a diverse benchmark suite to test single- and multi-agent \ac{rl} algorithms under various network conditions.
    \item Evaluation of single-agent baseline approaches as a foundation for future research.
\end{itemize}

Our implementation is available at  \href{https://github.com/jw3il/fairstream}{https://github.com/jw3il/fairstream}.
The remainder of this paper is structured as follows.
Sec.~\ref{sec:scenarios} presents the scenario of heterogeneous multimedia streaming and outlines associated challenges.
The following Sec.~\ref{sec:problem-statement} provides a formalization of the problem.
Sec.~\ref{sec:fair-streaming-env} introduces the streaming environment, including its configuration.
The experiments in Sec.~\ref{sec:experiments} show how the environment allows to analyze the behavior of agents.
We discuss the results in Sec.~\ref{sec:discussion}.
Sec.~\ref{sec:related-work} summarizes the related work and Sec.~\ref{sec:conclusion} concludes the paper.

\section{Scenario and Challenges}
\label{sec:scenarios}

\begin{figure*}[!t]
    \centering
    \includegraphics[width=\linewidth]{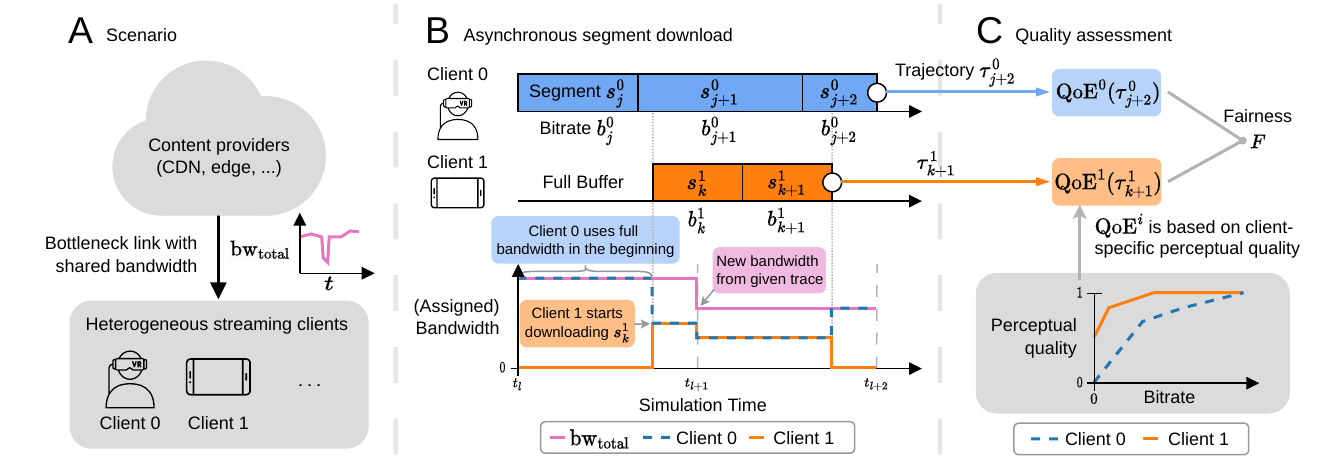}
    \caption{Streaming scenario with two exemplary clients. Subfigure A shows that all clients share a time-varying bottleneck link. Subfigure B depicts the asynchronous download of segments. Each rectangle represents a segment $s^i_t$ with bitrate $b^i_t$. The bottom graph shows that the total bandwidth $\text{bw}_\text{total}$ of the bottleneck is shared across all downloading clients. Subfigure C shows that the \ac{qoe} of each client depends on a client-specific function that maps the bitrate of a segment to a perceptual quality. To compute the fairness, the \ac{qoe} of all streaming clients is considered.}
    \label{fig:scenario}
    \vspace{-0.3cm}
\end{figure*}

The main goal of our streaming environment is to provide a benchmark environment for selected challenges in \ac{marl} research and to improve comparability in the area of fair multimedia streaming.
Each agent in the environment controls a streaming client and aims to maximize its quality while considering fairness across all clients. 
An exemplary scenario is shown in Fig.~\ref{fig:scenario}.
The following sections describe this scenario by highlighting the main emerging challenges.

\subsection{Partial Observability}
We consider heterogeneous streaming clients that access services from different content providers over a shared bottleneck link with a time-varying bandwidth (see Fig.~\ref{fig:scenario} A).
While most related works assume centralized control~\cite{nathan19minerva, yuan24qoeFairHeterogeneousMobile, subhan24jointQoEdrl}, we consider decentralized clients which make decisions based on locally available observations~\cite{seufert19tcpal}.
This includes application-specific observations such as the current buffer level, and network-specific observations such as previous download rates.
The clients do not receive any information about other clients within their observation and they have no information about future bandwidths.

This is challenging, as the dynamics of the streaming system depend on unknown and variable network conditions~\cite{stohr16QoEAnalysis}. 
Because of the shared bottleneck link, the actions taken by one client will further affect the network conditions observed by all other clients.
It has to be investigated to which degree and under which assumptions fairness can be reached with fully decentralized clients, and under which conditions communication between these clients becomes necessary.


\subsection{Agent Asynchronicity}
Clients in real communication systems usually act asynchronously in an event-based manner.
In our streaming scenario, agents select a new bitrate each time the assigned streaming client finishes downloading a segment (see Fig.~\ref{fig:scenario} B).
The time between steps varies, as the download duration of a segment depends on the available bandwidth and the chosen bitrate.
Additionally, there are no constraints on the ordering of the clients or the frequency of the steps.
One client might advance multiple steps during a single step of another client.

This is challenging, because the agents may act at different time scales. 
In contrast, existing \ac{marl} approaches~\cite{xueqiang24quic, saaid20dashfair} typically assume synchronous and equidistant steps. 

\subsection{Agent Heterogeneity}
Clients in streaming applications usually have different requirements based on the content type, level of user interactivity, device type and display resolution (see clients 0 and 1 in Fig.~\ref{fig:scenario} A).
These factors strongly influence the users' quality perception when interacting with the content~\cite{li16vmaf} and should be considered by \ac{abr} algorithms~\cite{seufert19tcpal}.
For example, mobile users might not notice artifacts that are obvious when viewing the same content on a television or with a \acl{hmd}.
To achieve the same \ac{qoe}, mobile users might be satisfied with lower bitrates.
%
For our scenario, we assume that the perceptual quality of a client is a function of the stream's bitrate $q^i(b)$ (see  Fig.~\ref{fig:scenario} C), similar to the work by \citet{mao17Pensieve} for homogeneous clients. 
In the heterogeneous case, however, the functions may differ between the clients~\cite{georgopoulos13openFlowQoEFair,subhan24jointQoEdrl}.
Higher bitrates are associated with higher perceptual qualities, but will lead to rebuffering if the total demanded bitrate of all clients exceeds the bandwidth of a shared bottleneck link.

This is challenging, as the reward functions differ between clients and the corresponding agents have to learn different behavior.


\subsection{Multiple Objectives}
The objective of our fair streaming scenario consists of two factors, a \ac{qoe} metric $\text{QoE}^i$ for each client $i \in I$ and a fairness metric $F$ over all clients (see Fig.~\ref{fig:scenario} C), similar to previous works that consider \ac{qoe} fairness \cite{saaid20dashfair, seufert19tcpal, nathan19minerva}.
Each client wants to maximize its own \ac{qoe}, while ensuring fairness over all clients.
We assume that a given bandwidth $\text{bw}_\text{total}$ of a shared bottleneck restricts the bitrates that can be chosen by the clients without leading to rebuffering, and therefore to a significant deterioration of the perceived quality of each individual client.

This is challenging, because with given bandwidth constraints, clients usually cannot maximize their own \ac{qoe} and achieve high fairness at the same time.
Instead, they have to trade off between the two components.

%


\section{Problem Statement}
\label{sec:problem-statement}
Considering these challenges, we now formulate the optimization problem associated with the streaming environment.
Its objective is composed of the individual \ac{qoe} of a client and the fairness across clients, which are detailed in the following subsections.
Finally, we propose a time-independent variant of the problem that will later be used to analyze the space of optimal solutions.

\subsection{Quality of Experience}
\label{sec:metrics-qoe}

Our platform allows researchers to integrate desired quality metrics for arbitrary content types. 
For reference, we define the \ac{qoe} of a streaming client based on the model by \citet{yin2015MPCStreaming}, which has been widely adopted by related works~\cite{mao17Pensieve, nathan19minerva}.
To keep the \ac{qoe} values bounded, we follow the suggestions of ITU-T Rec.~P.1203.3~\cite{ICU21QoEModel} and use an exponential decay to model the impact of the initial stalling delay and rebuffering during playback.
Under these considerations, we define the \ac{qoe} of streaming client $i$ at timestep $t\geq 0$ as

\begin{equation}
\label{eq:qoe}
\begin{aligned}
\text{QoE}^i\left(\tau^i_t\right) \coloneqq  &\underbracket[0pt][5pt]{\frac{q^i\left(b^i_t\right) + \delta \left[1 - \left| q^i\left(b^i_t\right) - q^i\left(b^i_{t-1}\right)\right| \right]_{t > 0}}{1 + \left[\delta\right]_{t > 0}}}_\text{First factor: normalized segment quality with switching penalty}  \\[5pt]
 &\,\cdot \underbracket[0pt][5pt]{\exp\left(-\lambda_\text{init} T_{\text{init}}\left(s^i_t\right) -\lambda_\text{reb} T_{\text{reb}}\left(s^i_t\right)\right)}_\text{Second factor: rebuffering penalty},
\end{aligned}
\end{equation}

where $\tau^i_t = (s^i_{-1}, b^i_0, s^i_0, \dots, b^i_t, s^i_t)$ is the trajectory of client $i$ after downloading segment $t \geq 0$. It contains the initial state $s^i_{-1}$ and all states $s^i_k$ after downloading segment $0 \leq k\leq t$ with bitrate $b^i_k \in \mathcal{B}^i$ from a finite set of available bitrates $\mathcal{B}^i$.

The first factor consists of the difference between the normalized perceptual quality $q^i(b^i_t) \in [0, 1]$ for bitrate $b^i_t$ with segment $t$ of agent $i$, and the switching penalty for the quality changes between the current and the previous segment $\left\vert q^i(b^i_t) - q^i(b^i_{t-1})\right\vert \in [0, 1]$, weighted by a coefficient $\delta \geq 0$. 
The switching penality for the initial step $t=0$ is zero, as indicated by the brackets.
We normalize the segment quality with the switching penality to $[0, 1]$ for improved interpretability.

The second factor represents the impact of rebuffering on the quality.
It is split into an initial buffering time $T_\text{init}(s) \in [0, \infty)$ when starting the stream and the rebuffering time during playback $T_\text{reb}(s) \in [0, \infty)$ with coefficients $\lambda_\text{init}, \lambda_\text{reb} \geq 0$.
Rebuffering significantly affects the perceived quality and should be avoided at all times by switching to lower bitrates when necessary.

Note that the perceptual quality $q^i(b_t^i)$ can be any function that represents the perceived quality for a given segment.
Examples include full-reference metrics like the \ac{ssim}~\cite{wang04SSIM} and fused quality assessment methods like VMAF \cite{li16vmaf}.
The same content can lead to different perceived qualities based on the media type, the device that is used for streaming, and the users' preferences, i.e., the perceptual quality of two clients $j \neq i$ may differ $q^i \neq q^j$.

\subsection{Fairness}
Fairness between clients can be interpreted in various ways.
Let $\vec{v} \coloneq (v^1,\,\dots,\,v^l) \in \mathbb{R}^l$ be a quality vector of $l$ clients for some quality metric.
Fairness measures map $\vec{v}$ to a score that represents the fairness of this solution. 
This work focuses on fairness in terms of \ac{qoe} between clients.

While notions of fairness such as Jain's fairness index \cite{jain1998quantitative} and max-min fairness are widely applied in the context of communication systems, the \ac{qoe} fairness index F by \citeauthor{hossfeld17QoEFairness} is specifically designed with \ac{qoe} in mind \cite{hossfeld17QoEFairness}:
\begin{equation}
F(\vec{v}) \coloneqq 1 -\frac{\sigma(\vec{v})}{\sigma_\text{max}} = 1 - \frac{2\sigma(\vec{v})}{H - L}.
\end{equation}

By normalizing the standard deviation $\sigma(\vec{v})$ of the qualities according to their range $v^k \in \left[L,\, H\right]$, $F$ is independent of the quality range. 
This is important when comparing \ac{qoe} definitions with different ranges.
\citet{kim22AdaptiveStreaming} follow a similar line of thought by normalizing \ac{qoe} differences in their utility function.
We integrate the normalization directly into the \ac{qoe} definition, see Eq.~(\ref{eq:qoe}).
The fairness index $F$ 
yields scores in $[0, 1]$, where higher values indicate higher fairness.
A value of $0$ is reached for the maximum standard deviation, i.e.\ when half of the clients retrieve scores $L$ and $H$, respectively.
A value of $1$ indicates that all clients have the same \ac{qoe}.

To capture the average streaming quality over time, we consider fairness over an exponential moving average~\cite{owen17ema} of the \ac{qoe} with smoothing factor $\kappa \in [0, 1]$
\begin{equation}
    \label{eq:qoe-ema}
    \begin{aligned}
        v^i_t \coloneqq \frac{z^i_t}{1 - \kappa^{t+1}}
        \quad\text{ where }\quad z^i_t = \begin{cases}
    0& \text{if } t = -1\\
    \kappa z^i_{t-1} + (1 - \kappa) \text{QoE}^i(\tau^i_t) & \text{otherwise}
\end{cases}
    \end{aligned}
\end{equation}

This only considers clients $i \in L_{T^i_t} \subseteq I$ that are streaming at simulation time $T^i_t \in \mathbb{R}$, denoting the elapsed time since starting the environment when client $i$ finished performing step $t\geq 0$.

The vector of exponentially averaged \ac{qoe} values at the time when client $i$ finishes downloading segment $t$ is given as 
\begin{equation}
    \vec{v^i_t} \coloneq \left(v^1_{\text{last}^1(T^i_t)},\,\dots,\,v^l_{\text{last}^l(T^i_t)}\right),
\end{equation}

where {\thickmuskip=0mu $\text{last}^k(T)$}$\: \in \mathbb{N}$ indicates the last completed step of client $k$ at simulation time $T\in \mathbb{R}$.
The moving average is bounded $\vec{v^i_t} \in [0, 1]^l$ and used as input for fairness index $F$.

\subsection{Utility Function}

The agents' goal is to maximize their \ac{qoe} and the fairness between clients.
We consider the naive linear combination
\begin{equation}
\begin{alignedat}{2}
    U^i_\alpha\left(\vec{\tau^i_t}\right) \coloneqq&\ \alpha \textit{QoE}^i\left(\tau^i_t\right) + (1 - \alpha) F\left(\vec{v^i_t}\right)\\
    =&\ \alpha \textit{QoE}^i\left(\tau^i_t\right) + (1 - \alpha)\left(1 - 2\sigma\left({\vec{v^i_t}}\right)\right)
\end{alignedat}
\label{eq:combined-objective-reward}
\end{equation}

for each client $i$ with a quality-fairness coefficient $\alpha \in [0, 1]$ and the trajectories $\vec{\tau^i_t} \coloneq \left(\tau^1_{\text{last}^1(T^i_t)}, \dots, \tau^l_{\text{last}^l(T^i_t)}\right)$ of all clients from the perspective of client $i$.
In this formulation, the optimal client behavior depends on the quality-fairness coefficient $\alpha$.
A high $\alpha \rightarrow 1$ leads to selfish clients that try to maximize their own \ac{qoe}, while $\alpha \rightarrow 0$ leads to clients that neglect their own quality to prioritize fairness.
In the following, we refer to this combined objective as \emph{utility} and will explore the effect of coefficient $\alpha$ based on a time-independent version of this problem.

\subsection{Time-Independent Formulation}
\label{sec:time-independent-formulation}

For a computationally tractable investigation of the effect of coefficient $\alpha$ on the space of optimal solutions, we propose a simplified version of the optimization problem with a time-independent bandwidth.
Instead of storing downloaded segments in a buffer, we assume that segments of infinitesimal size are played back instantly.
The number and type of clients are fixed for the whole duration and the stream continues indefinitely.
Under these assumptions, optimal policies with global knowledge simply choose the bitrate that leads to the highest utility at the beginning and stick to that choice.
As there can be no quality switches without the time dimension, the \ac{qoe} from Eq.~(\ref{eq:qoe}) reduces to the perceptual quality $q^i(b)$.
As the quality does not change over time, the exponential moving average $v^i_t$ of the \ac{qoe} from Eq.~(\ref{eq:qoe-ema}) also reduces to the perceptual quality.
The fairness is therefore defined as $F$ applied to the perceptual qualities of all clients.
As our goal is to maximize the combined objective from Eq.~(\ref{eq:combined-objective-reward}) over all clients, we consider its arithmetic average as the objective function.
As the clients' quality is static for a solution, the fairness is equal for all clients in this formulation.
Therefore, the objective consists of the average perceptual quality and the fairness over all clients.

We are only interested in solutions that would not feature any rebuffering if transferred to the original problem.
This is achieved with a constraint that ensures that the total bitrate of all clients does not exceed a given total bandwidth $\text{bw}_{\text{total}}$, allowing for seamless playback. 

The resulting optimization problem for the time-independent case is summarized in Eq.~\ref{eq:optim}:

\begin{equation}
\begin{alignedat}{2}
&\!\max_{b}        &\qquad& \alpha \left(\frac{1}{\lvert I \rvert}\sum_{i\in I} q^i(b^i)\right) \\
&       & \qquad &  + \left(1 - \alpha\right) \left(1 - 2\sigma\left(\left(q^i\left(b^i\right)\right)_{i\in I}\right)\right)\\[8pt]
&\text{subject to} &     & \sum_{i \in I} b^i \leq  \text{bw}_{\text{total}} ,\\
&                  &      & b^i \in \mathcal{B}^i \quad  \forall i \in I.
\end{alignedat}
\label{eq:optim}
\end{equation}

Solving this problem is nontrivial, as the objective function is nonlinear and the solution space scales exponentially with the number of clients $\lvert I \rvert$ and the number of considered bitrates  $\lvert\mathcal{B}^i\rvert$.
However, when considering few clients and a small number of bitrates, it can easily be solved by full enumeration over all bitrates.
We analyze the space of optimal solutions in Sec.~\ref{sec:quality-fairness-coefficient-choice}.

\section{Environment and Experiment Design}
\label{sec:fair-streaming-env}
In the following, we build upon the problem statement and describe the streaming environment.

\subsection{Streaming Clients}
\label{sec:env-clients}
\begin{figure}[!h]
    \centering
    \includegraphics[width=0.45\linewidth]{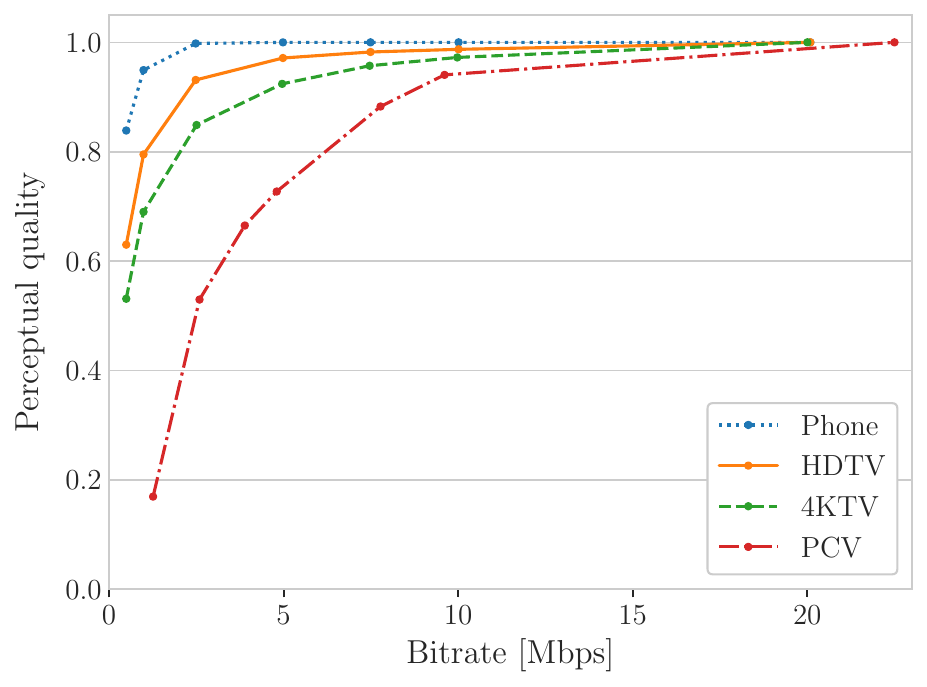}
    \vspace{-0.2cm}
    \caption{Bitrates and corresponding perceptual qualities for the four considered client types. The different slopes indicate different resource requirements.}
    \label{fig:client-quality-overview}
\end{figure}

In our streaming environment, each agent controls a multimedia client. 
We consider four exemplary client types.
Their individual perceptual quality functions $q^i(b)$ are shown in Fig.~\ref{fig:client-quality-overview}.
The first three client types represent streaming of traditional video on a phone (Phone), a HD television (HDTV), and a 4K television (4KTV).
The mapping from bitrates to perceptual qualities is based on  results of the \ac{vmaf} model~\cite{li16vmaf} for the Big Buck Bunny movie~\cite{bbb3d}.
The last client represents streaming a \ac{pcv} with normalized quality values taken from a subjective study~\cite{weil2023PointCloudQoEModelling}.
We select seven quality settings for each client type.
Further details are provided in the appendix, see Sec.~\ref{appendix:perceptual-quality}.

All clients reach higher quality levels for higher bitrates, but the increase in quality per bit depends on the client type.
For example, the PCV stream requires more than 7.5 Mbps to reach a quality that is comparable with the lowest quality setting of the Phone stream at around 0.5 Mbps.

Each agent controls a client and has a discrete action space according to the available bitrates $\mathcal{B}^i$. 
Based on the considered quality settings, each agent has $\lvert\mathcal{B}^i\rvert = 7$ actions.
In our case, the minimum total bandwidth required for all clients to stream seamlessly with the lowest quality is $2.75$ Mbps. 
For all clients to stream with the highest quality, a total bandwidth of $82.68$ Mbps is required.

\subsection{Bandwidth of Bottleneck Link}
\label{sec:env-traces}
We assume that the bandwidth of the bottleneck link varies over time.
Consequently, the agents have to estimate the available bandwidth in order to select bitrates that yield the highest utility.
The simulations should cover a variety of different scenarios.
For example, there should be scenarios with a limited bandwidth that is not sufficient to stream with higher quality settings.
Scenarios with bandwidth fluctuations are also relevant, e.g., caused by unstable connections.

\begin{figure}[!t]
    \centering
    \begin{tikzpicture}
    \node[inner sep=0pt] (a) {\includegraphics[height=3.3cm]{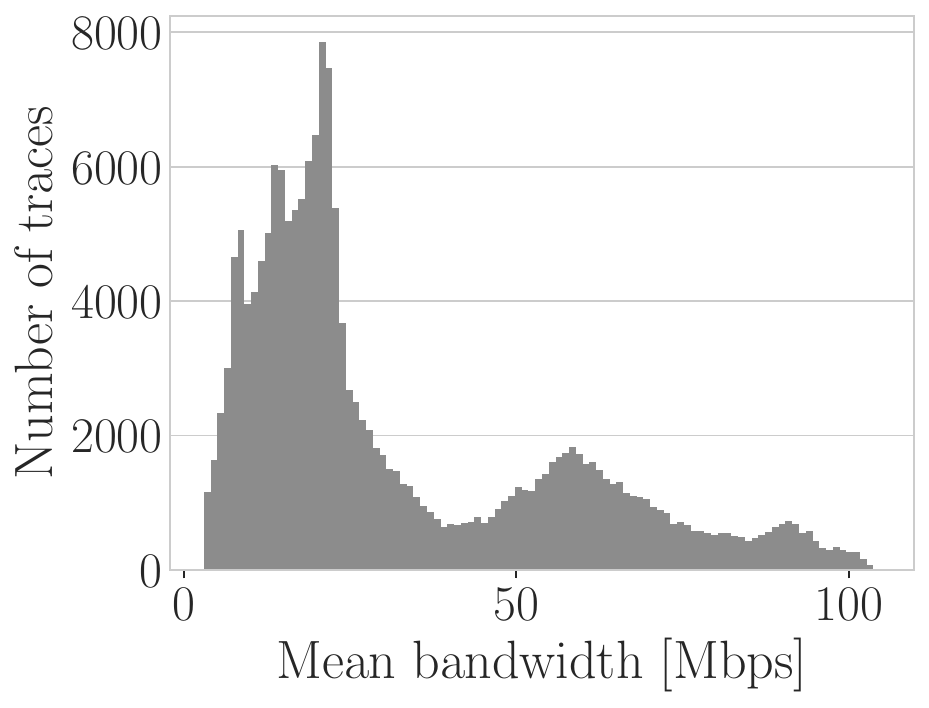}};
    \node[above, inner sep=0pt] at ([shift={(0.2265cm,0.1cm)}] a.north) {a) Bandwidth of all traces};
    \node[inner sep=0pt] (b) [right= 0.25cm of a] {\includegraphics[height=3.3cm]{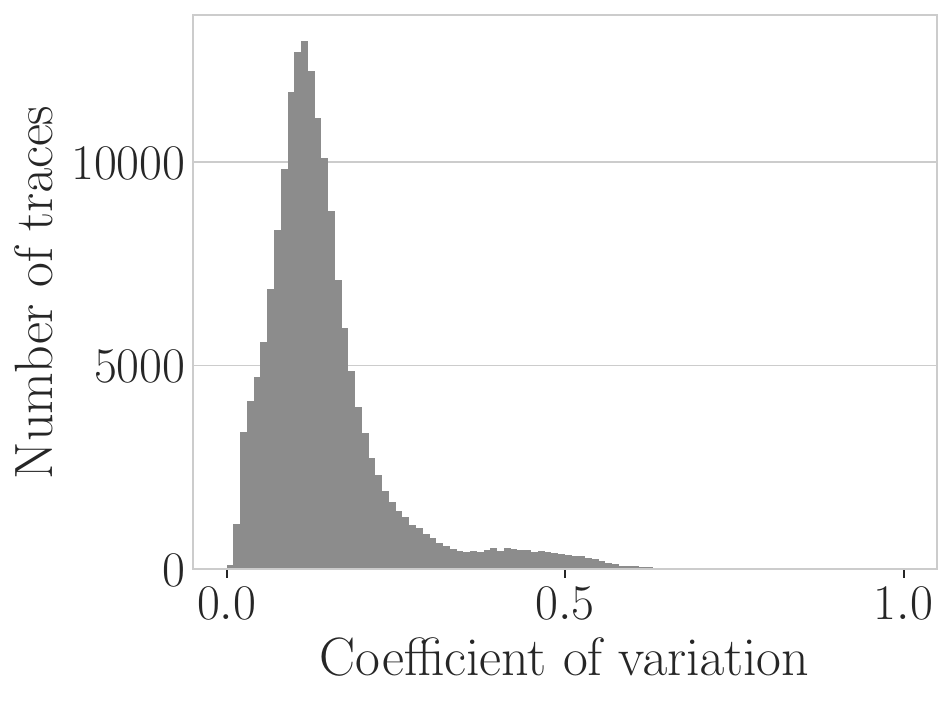}};
    \node[above, inner sep=0pt] at ([shift={(0.375cm,0.1cm)}]b.north) {b) \acs{cv} of all traces};
    \node[inner sep=0pt] (c) [right=0.2cm of b] {\includegraphics[height=3.3cm]{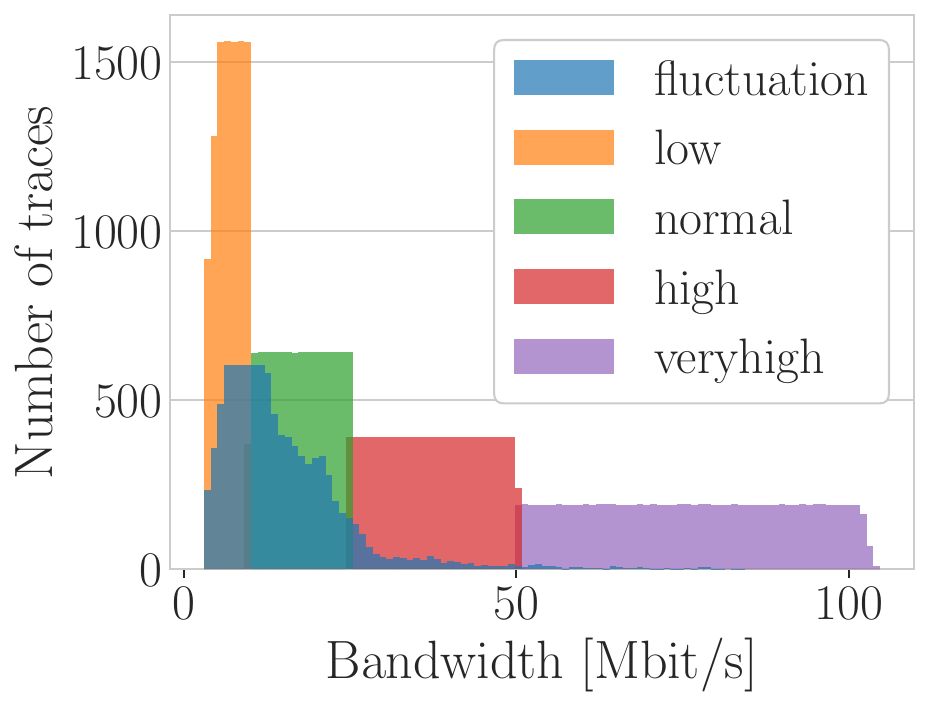}};
    \node[above, inner sep=0pt] at ([shift={(0.265cm,0.1cm)}]c.north) {c) Extracted trace dataset};
    \end{tikzpicture}
    \caption{Mean bandwidth a) and \acf{cv} b) of all traces, as well as the mean bandwidth of the traces of our dataset c). Subplot b) in the center shows traces with a \ac{cv} in $[0, 1]$, representing $99.6\%$ of all data. Traces with a \ac{cv} greater than $1$ are very infrequent and would not be visible in this histogram.}
    \label{fig:all-traces}
\end{figure}
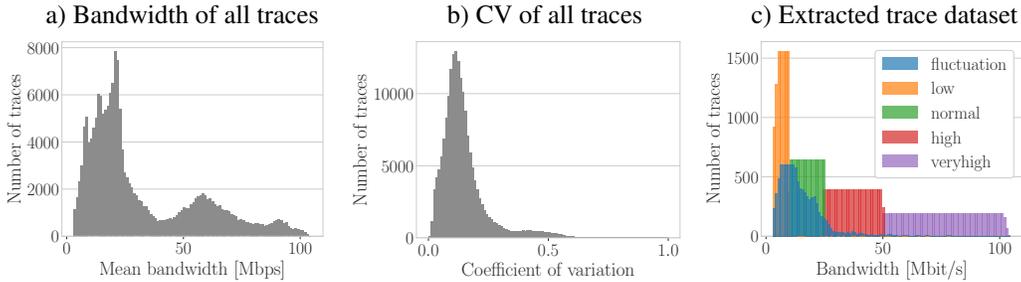

The Federal Communications Commission (FCC) provides bandwidth measurements of American households' broadband connections based on HTTP requests to popular web pages~\cite{fccDatasets}.
We extract traces of $200$ seconds from May 2022 to July 2023 by concatenating and cutting all measurements between unique clients and destinations.
Finally, we scale the throughput by a factor of three to fit the bandwidth requirements of our four client types.
Filtering all traces with invalid values and a mean bandwidth below 3 Mbit/s yields a total of $176\:873$ traces which are illustrated in Fig.~\ref{fig:all-traces}.
Note that the traces still contain measurements under 3 Mbit/s, but should allow for seamless playback on the lowest quality setting on average.
The \ac{cv} shows the standard deviation of the bandwidth divided by the average bandwidth.
High values represent varying bandwidths, e.g.,~transitions between low and high bandwidths within the trace. 

The original trace distribution shows three modes at around 25 Mbps, 60 Mbps and 90 Mbps.
Training and evaluating agents on the original trace distribution comes with the risk of introducing a strong bias towards traces around 25 Mbps, as these occur most frequently.
Additionally, the comparatively low number of fluctuating traces with a \ac{cv} above $0.35$ would very likely cause agents to ignore such cases.
We therefore undersample the original traces based on their \ac{cv} and average bandwidth.
Traces with a \ac{cv} greater than $0.35$ are assigned to the \emph{fluctuating} class, all remaining traces are classified as \emph{low}, \emph{normal}, \emph{high}, and \emph{veryhigh} according to their average bandwidth.

\begin{table}[!t]
\caption{Network trace dataset with classes according to the \acl{cv} (\acs{cv}) and average bandwidth.}
    \small
\begin{tabular}{lllll}
\toprule
\textbf{Traffic class} & \textbf{CV} & \multicolumn{2}{l}{\textbf{Average bandwidth}} & \textbf{\# Traces}\\[3pt]
 & & lower $(>)$ & upper $(\leq)$ & \\ \midrule
fluctuating & $\geq 0.35$ & 3 Mbps & $\infty$ & $10\:000$ \\ 
low & $< 0.35$ & 3 Mbps & 10 Mbps & $10\:000$ \\ 
normal & $< 0.35$ & 10 Mbps & 25 Mbps & $10\:000$ \\ 
high & $< 0.35$ & 25 Mbps & 50 Mbps & $10\:000$ \\ 
veryhigh & $< 0.35$ & 50 Mbps & $\infty$ & $10\:000$ \\\bottomrule
\end{tabular}
\centering
\label{tab:dataset}
\end{table}

We randomly sample $10\:000$ network traces for each class so that the mean bandwidth per class is approximately uniformely distributed, see Fig.~\ref{fig:all-traces}. The limits are summarized in Table~\ref{tab:dataset}.
With $200$ seconds of simulation time per trace, the $50\:000$ traces allow for over $2\:000$ hours of simulations.

The traces are split into three sets: $90\%$~of the traces are used for training, $5\%$ for validation, and $5\%$ for testing.
The traces are passed to the environment upon initialization.
Specific network conditions could easily be simulated by modifying how traces are sampled from these sets.
For example, the fluctuating traces can be left out to simulate environments with comparatively stable network conditions.

\subsection{Bandwidth Allocation}
\label{sec:env-bandwidth}
Motivated by the bandwidth slicing approach of \citet{nathan19minerva}, we consider a \emph{weight-based} bandwidth allocation mechanism.
At simulation time $T$, the total bandwidth of the bottleneck link $\text{bw}_{\text{total}}(T)$ is distributed across all clients that are downloading segments $D_T \subseteq I$ according to
\begin{equation}
    \text{bw}^i_{\text{weighted}}(T) := \text{bw}_{\text{total}}(T) \, \frac{w^i}{\sum_{j\in D_T}w^j},\: i \in D_T,
\end{equation}
where $w^k$ is the weight of client $k\in D_T$.

Note that this covers static and dynamic bandwidth allocation schemes.
For example, fixing $w^i_t = 1$ for all clients would result in equal bandwidth allocation irrespective of the demands of each client, similar to the concept of TCP fairness.
In comparison, setting $w^i_t = b^i_t$ results in a bandwidth allocation that is proportional to the requested bitrates $b^i_t$. 
In this case, segments of equal duration would require the same time to download, irrespective of the individual bitrates.

\subsection{Observations}
\label{sec:env-observation}
At the beginning of an episode and after downloading a segment $s^i_t$ at step $t$, agent $i$ receives a partial observation $o^i_t$ representing the view of the client.
This observation is of form

\begin{equation}
\begin{aligned}
o^i_t \coloneqq \big(&\text{QoE}^i(\tau^i_t),~ v^i_t,~q^i_t(b^i_t),~ b^i_t,~\Delta T^i_t,\\&T_\text{init} (s^i_t),~T_\text{reb} (s^i_t),~\text{bf}^i_t,~c^i_t,~\vec{b^i},~\vec{q^i}\big)
\end{aligned}
\end{equation}

with the following elements

\begin{itemize}[leftmargin=2.5cm]
\item[$\text{QoE}^i(\tau^i_t)$] The \ac{qoe} associated with this segment. This component  is directly connected to the reward of the agents. Agents should learn to increase their \ac{qoe} by adapting their bitrate depending on the network conditions.\
\item[$v^i_t$] The exponential moving average of the \ac{qoe}, as considered in the fairness metric. Note that the fairness is not included in the agent's observations, because it requires global knowledge.\
\item[$q^i_t(b^i_t)$] The perceptual quality of the segment.\ 
It is the main contributor of the \ac{qoe}. Agents should learn to increase their perceptual quality in an under-utilized network.\
\item[$b^i_t$] The bitrate of the downloaded segment.\
\item[$\Delta T^i_t$] The time spent downloading the last segment $\Delta T^i_t \coloneqq T^i_t - T^i_{t-1}$.
Especially in the beginning, it is crucial to keep the download time lower than the segment time to avoid re-buffering.
Short download durations suggest that a higher bitrate can be chosen.\
\item[$T_\text{init} (s^i_t)$] The initial stalling time while downloading the last segment, only happens at the beginning of the stream.\
\item[$T_\text{reb} (s^i_t)$] The time spent rebuffering while downloading the last segment.\
\item[$\text{bf\,}^i_t$] The current buffer level of the client. The risk of stalling by choosing a higher bit rate decreases with a higher buffer level. A high buffer level indicates that it is safe to switch to a higher bitrate to increase the client's \ac{qoe}.
\item[$c^i_t$] The number of remaining segments before the stream ends.\
\item[$\vec{b^i}$] Vector of selectable bitrates $\vec{b^i} \coloneq \left((b)\right)_{b \in B^i}$.\
\item[$\vec{q^i}$] Vector of perceptual qualities $\vec{q^i} \coloneq \left(q^i(b)\right)_{b \in B^i}$.
\end{itemize}

Note that the clients receive no information about other clients in the network by default.
Custom agents may augment the observation space, e.g., by adding a communication channel between agents that allows them to coordinate.

\subsection{Optimal Solutions of the Time-Independent Formulation}
\label{sec:quality-fairness-coefficient-choice}

We expect the choice of the quality-fairness coefficient $\alpha$ from Eq.~(\ref{eq:combined-objective-reward}) to strongly affect the optimal agent behavior.
In order to make a well-founded selection of $\alpha$ for the reward function, we first analyze the space of optimal solutions of the time-independent formulation in Eq.~(\ref{eq:optim}) from Sec.~\ref{sec:time-independent-formulation}.

\begin{figure}[!b]
    \centering    \includegraphics[width=0.55\linewidth]{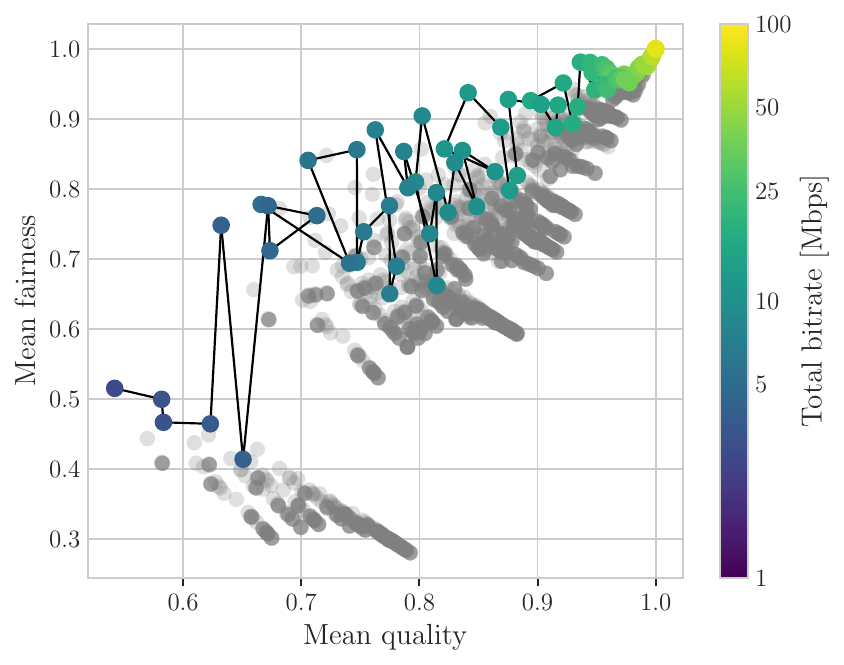}
    \caption{Overview of all feasible solutions (grey transparent) and pareto-optimal solutions (colored) of the time-independent formulation. Optimal solutions are connected by a line according to their ordered bitrate.}
    \label{fig:pareto}
\end{figure}

Fig.~\ref{fig:pareto} provides an overview of the feasible and pareto-optimal solutions of the problem, using the total bitrate of a solution as the bandwidth constraint.
The feasible solutions represent all bitrate combinations that can be selected by clients.
With pareto-optimal solutions, we denote feasible solutions that are not dominated by other solutions with a lower or equal total bitrate, i.e.~there are no solutions with less resource requirements that have the same or higher quality and fairness and a higher sum of both metrics.
Depending on the parameter $\alpha$ and a given bandwidth $\text{bw}_{\text{total}}$, the optimal solutions of Eq.~(\ref{eq:optim}) are a subset of these pareto-optimal solutions.
The plot suggests that bandwidths below $25$ Mbps will be the most challenging for learning-based approaches, as the pareto-optimal solutions differ greatly and small changes in an agent's actions might have a big effect on its return.
Above $25$ Mbps, the solutions are quickly getting close to the optimum of $(1, 1)$ with smaller steps.
This should be easier to learn.

\begin{figure}[!t]
    \centering
    \begin{tikzpicture}
	\newcommand{\topologyfigheight}{6.5cm}
	\node[inner sep=0pt] (a) {\includegraphics[height=\topologyfigheight,trim={0 0 0 0},clip]{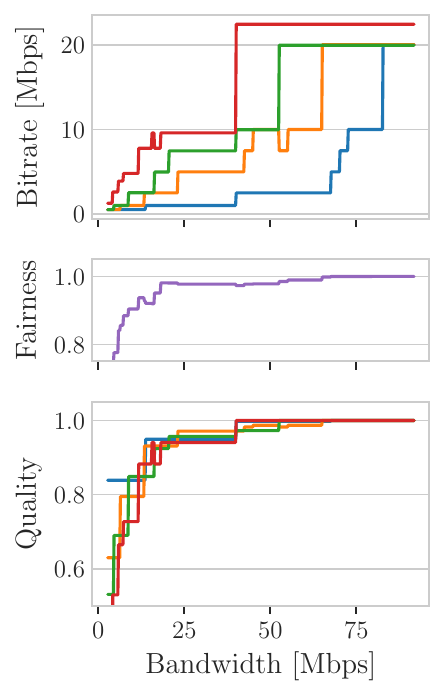}};
	\node[inner sep=0pt] [above right= 0.0cm and -2.2cm of a]{$\alpha=0.25$};
	\node[inner sep=0pt] (b) [right= 0.5cm of a] {\includegraphics[height=\topologyfigheight,trim={1.44cm 0 0 0},clip]{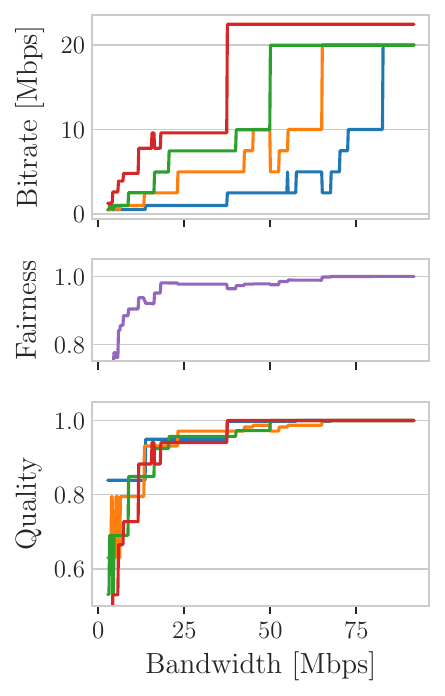}};
	\node[inner sep=0pt] [above=0.0cm of b]{$\alpha=0.5$};
	\node[inner sep=0pt] (c) [right= 0.5cm of b] {\includegraphics[height=\topologyfigheight,trim={1.44cm 0 0 0},clip]{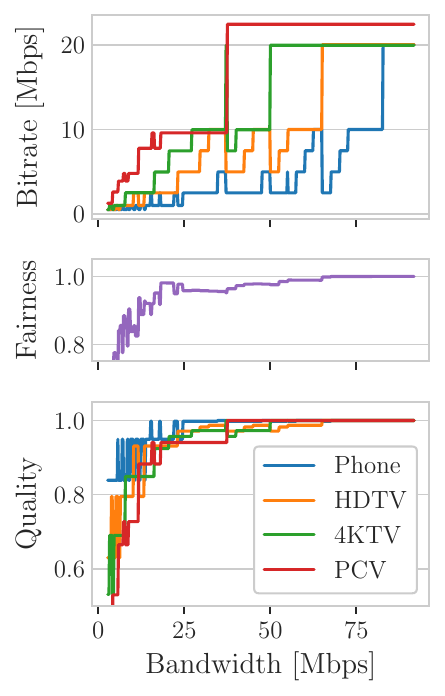}};
	\node[inner sep=0pt] [above=0.0cm of c]{$\alpha=0.75$};
    \end{tikzpicture}
    \caption{Optimal solutions for the time-independent formulation with four clients Phone, HDTV, 4KTV and PCV using different quality-fairness coefficients $\alpha = 0.25$ (left), $\alpha = 0.5$ (center), and $\alpha = 0.75$ (right). The top plots show the bitrate of each client, given the bandwidth according to the horizontal axis. The fairness between all qualities is depicted in the center. The bottom plots show the quality of each client. 
    For higher $\alpha$, clients prioritize quality over fairness at the cost of more frequent bitrate and quality changes.}
    \label{fig:alpha-comparison}
\end{figure}
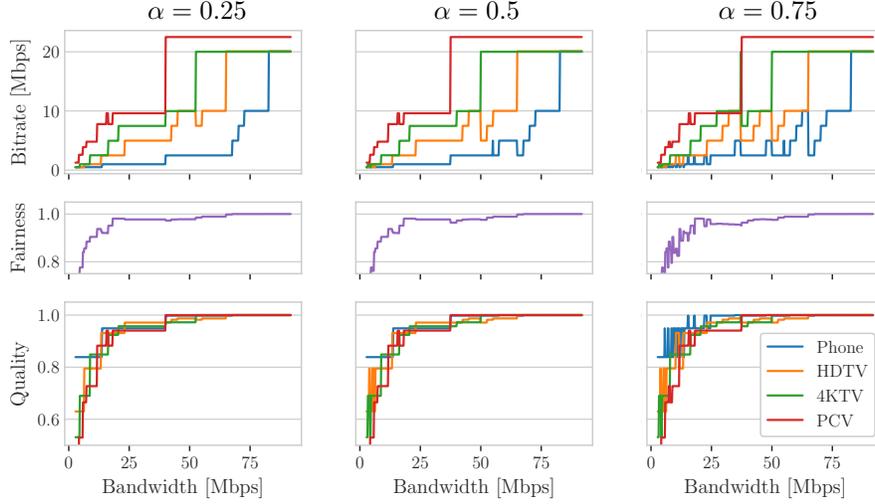
\begin{figure}[!t]
    \centering
    \begin{tikzpicture}
	\newcommand{\topologyfigheight}{4.7cm}
	\node[inner sep=0pt] (a) {\includegraphics[height=\topologyfigheight,trim={0 0 0 0},clip]{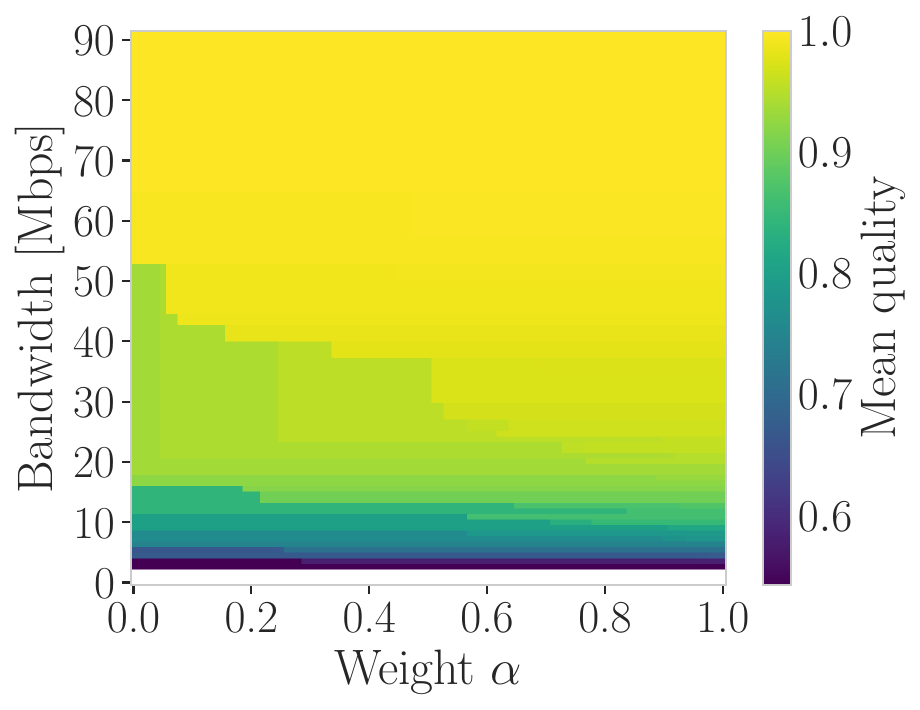}};
	\node[inner sep=0pt] [above right= 0.0cm and -4.4cm of a]{a) Mean quality};
	\node[inner sep=0pt] (b) [right= 0.5cm of a] {\includegraphics[height=\topologyfigheight,trim={0 0 0 0},clip]{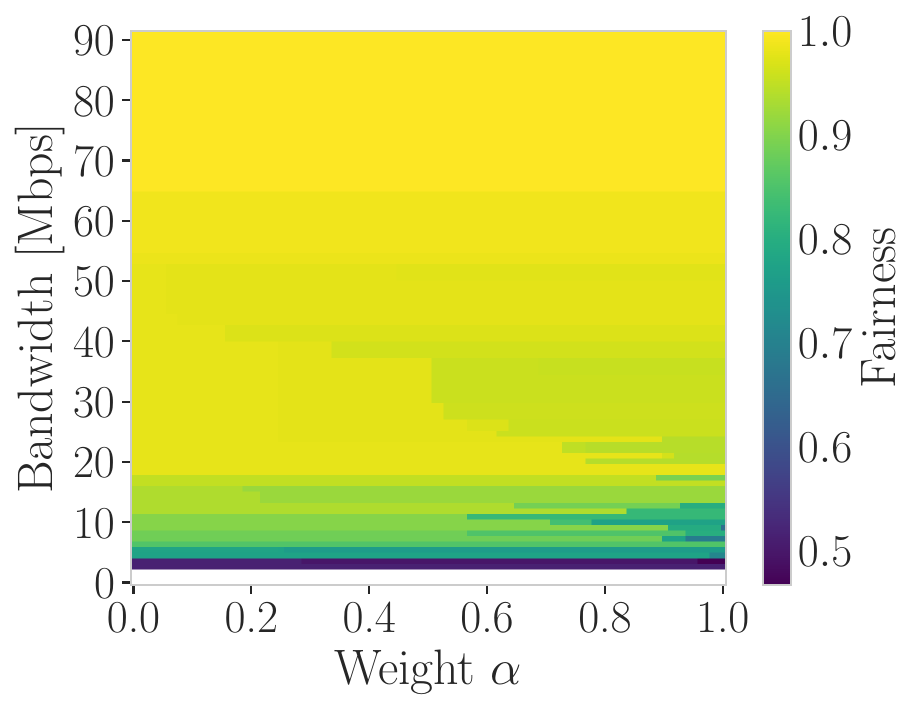}};
	\node[inner sep=0pt] [above right= 0.0cm and -4cm of b]{b) Fairness};
	\draw[draw=black] (b)++(0.75,-0.25) rectangle ++(1.0,-1.1);
    \end{tikzpicture}
    \caption{Mean quality a) and fairness $F$ b) of the optimal solutions for the time-independent formulation with four heterogeneous clients using $\alpha \in [0,\,1]$ and bandwidths in $[0, 90]$ Mbps. The white horizontal line at the bottom indicates that there is insufficient bandwidth for seamless playback. Values of $\alpha > 0.5$ lead to higher quality at the cost of deteriorated fairness. The black rectangle in the right figure b) highlights that the fairness of the optimal solutions does not always increase with higher bandwidth.}
    \label{fig:alpha-all}
\end{figure}
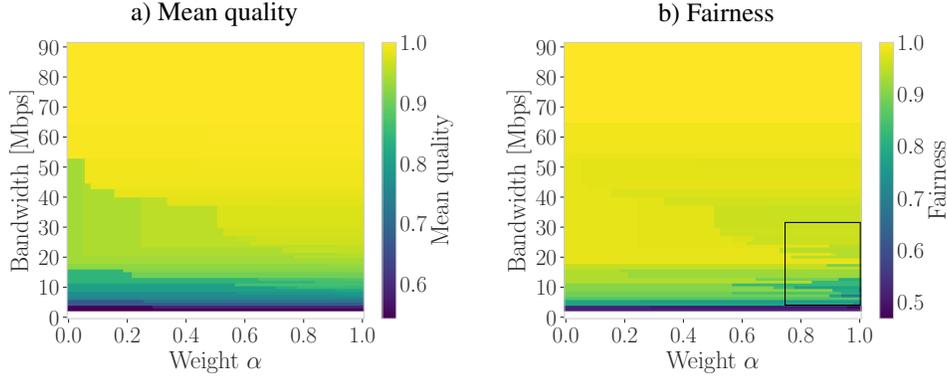

Fig.~\ref{fig:alpha-comparison} shows the optimal solutions for the clients from Fig.~\ref{fig:client-quality-overview} for different bandwidths $\text{bw}_{\text{total}}$ and $\alpha \in \{0.25,\, 0.5,\, 0.75\}$.
Remember that the quality-fairness coefficient $\alpha$ balances quality and fairness in the objective.
A higher $\alpha$ means that more weight is assigned to the \ac{qoe}.
The figure indicates that this leads to more frequent and earlier bitrate changes, while the difference between the qualities increases.
Although this leads to a higher mean quality, these rapid quality changes would be undesirable with respect to the stability of the learning target and result in lower fairness.

The effect of coefficient $\alpha \in [0, 1]$ on the mean quality and fairness is visualized in Fig.~\ref{fig:alpha-all}.
Each pixel in the figure represents an optimal solution for one instance of the optimization problem.
The bandwidth and $\alpha$ were both sampled with 100 uniform steps in the shown ranges.
Note that the shape of the blocks in the figure originate from the optimization problem itself, they are not caused by the visualization.
Similar to the previous findings, we can see that the optimal solutions show more fine-grained quality steps for higher $\alpha > 0.5$.
At the same time, the fairness of these solutions drops significantly.
Overall, the optimal solutions show comparatively high mean quality and fairness for all bandwidths greater than 25 Mbps.

\subsection{Reward Function}
\label{sec:reward-function}
Based on the results from  Sec.~\ref{sec:quality-fairness-coefficient-choice}, we set $\alpha = 0.25$ for the majority of our experiments to reduce the possibility that fairness gets overshadowed by the client's individual \ac{qoe}.
The switching penality coefficient is set to $\delta = 0.025$ analogous to the paramaters chosen by \citet{nathan19minerva}.
The rebuffering penalty coefficients are $\lambda_{\text{init}} = 1$ and $\lambda_{\text{reb}} = 10$, with the intention to greatly reduce the \ac{qoe} during playback upon rebuffering. For example, rebuffering for $0.1$ seconds reduces \ac{qoe} by around $63\%$.
The resulting reward $R^i_t$ of agent $i$ after selecting bitrate $b^i_t$ at at step $t$ is given by

\begin{equation}
\begin{alignedat}{2}
    R^i_t \coloneqq U^i_{0.25}\left(\vec{\tau^i_t}\right).
\end{alignedat}
\label{eq:chosen-reward}
\end{equation}

In our experiments, we also provide a brief comparison with $\alpha = 0.5$ and $\alpha = 0.75$.
Future research could investigate vector-based rewards with independent scalar values for each objective or even for each individual component of the \ac{qoe} function.

\section{Results}
\label{sec:experiments}

We implement the environment based on the RLlib multi-agent interface~\cite{liang2018rllib}
and consider the following agent types:
\begin{itemize}
    \item \emph{Min}: Always chooses the minimum bitrate.
    \item \emph{Max}: Always chooses the maximum bitrate.
    \item \emph{Random}: Chooses a random bitrate at each step.
    \item \emph{Greedy-$k$}: Heuristic baseline agent that was implemented for this environment. The agent initially selects the minimum bitrate. Then, it computes the average download rate of up to $k$ previous segments. Assuming this is the available bandwidth of future steps, the agent greedily selects the maximum bitrate that could be streamed without rebuffering.
    \item \emph{PPO}: Agents trained using the \ac{ppo} algorithm~\cite{schulman2017PPO}. \ac{ppo} is considered one of the state-of-the-art \ac{rl} algorithms~\cite{yu22ppoInMultiAgent} and has shown to outperform previous approaches when learning \ac{abr} control~\cite{naresh23ppoABR}.
\end{itemize}

For bandwidth sharing, we consider two modes:
\begin{itemize}
    \item \emph{Proportional}: The bandwidth is distributed proportionally to the selected bitrate of each client.
    \item \emph{Minerva}: The Minerva algorithm by~\citet{nathan19minerva} considers a linear interpolation between the discrete bitrate-quality mappings of each client and computes the bandwidth shares that would allow each agent to stream at the same interpolated quality.
\end{itemize}


The following Sec.~\ref{sec:eval-overview} provides an overview of the results for all approaches.
In Sec.~\ref{sec:eval-training}, we shows the results of the validation runs performed during training.
The following Sec.~\ref{sec:eval-clients} shows how the performance metrics vary between the heterogeneous clients.
Sec.~\ref{sec:eval-trace} details how agents act based on an exemplary trace and Sec.~\ref{sec:ppo-effect-fairness-coeff} shows the effect of different values for the quality-fairness coefficient on the \ac{ppo} agent.

\begin{table}[!t]
    \centering
    \caption{Results on the test traces aggregated over all clients and traffic classes. The first row represent the agents and the following rows results for individual metrics. The G8 agent is an abbreviation for Greedy-$8$. Agents with the suffix ``-Minerva'' use Minerva bandwidth sharing, the others use proportional bandwidth sharing.}
    \resizebox{\linewidth}{!}{%
    \begin{tabular}{lccccccc}
\toprule
\textbf{Metric} / \textbf{Agent}  & Min & Max & Random & G8 & G8-Minerva & PPO & PPO-Minerva\\ \midrule
Return $\uparrow$ & $55.15\, {\scriptstyle \pm\, 6.00}$  & $44.87\, {\scriptstyle \pm\, 29.53}$  & $67.07\, {\scriptstyle \pm\, 19.38}$  & $85.15\, {\scriptstyle \pm\, 12.57}$  & $89.32\, {\scriptstyle \pm\, 9.78}$  & $88.93\, {\scriptstyle \pm\, 8.74}$  & $81.73\, {\scriptstyle \pm\, 12.24}$ \\
\ac{qoe} $\uparrow$ & $0.55\, {\scriptstyle \pm\, 0.23}$  & $0.11\, {\scriptstyle \pm\, 0.27}$  & $0.47\, {\scriptstyle \pm\, 0.35}$  & $0.85\, {\scriptstyle \pm\, 0.16}$  & $0.85\, {\scriptstyle \pm\, 0.16}$  & $0.69\, {\scriptstyle \pm\, 0.32}$  & $0.74\, {\scriptstyle \pm\, 0.17}$ \\
Fairness $\uparrow$ & $0.55\, {\scriptstyle \pm\, 0.01}$  & $1.00\, {\scriptstyle \pm\, 0.00}$  & $0.88\, {\scriptstyle \pm\, 0.08}$  & $0.85\, {\scriptstyle \pm\, 0.12}$  & $0.90\, {\scriptstyle \pm\, 0.09}$  & $0.96\, {\scriptstyle \pm\, 0.02}$  & $0.83\, {\scriptstyle \pm\, 0.10}$ \\
Perceptual Quality $\uparrow$ & $0.54\, {\scriptstyle \pm\, 0.24}$  & $1.00\, {\scriptstyle \pm\, 0.00}$  & $0.85\, {\scriptstyle \pm\, 0.10}$  & $0.86\, {\scriptstyle \pm\, 0.16}$  & $0.87\, {\scriptstyle \pm\, 0.12}$  & $0.90\, {\scriptstyle \pm\, 0.07}$  & $0.82\, {\scriptstyle \pm\, 0.08}$ \\
Init Rebuffer Time [s] $\downarrow$ & $0.21\, {\scriptstyle \pm\, 0.25}$  & $5.57\, {\scriptstyle \pm\, 5.31}$  & $1.84\, {\scriptstyle \pm\, 2.19}$  & $0.21\, {\scriptstyle \pm\, 0.25}$  & $0.32\, {\scriptstyle \pm\, 0.44}$  & $0.64\, {\scriptstyle \pm\, 0.69}$  & $0.98\, {\scriptstyle \pm\, 3.14}$ \\
Rebuffer Time [s] $\downarrow$ & $0.00\, {\scriptstyle \pm\, 0.05}$  & $4.69\, {\scriptstyle \pm\, 5.01}$  & $1.12\, {\scriptstyle \pm\, 1.60}$  & $0.01\, {\scriptstyle \pm\, 0.06}$  & $0.02\, {\scriptstyle \pm\, 0.12}$  & $0.21\, {\scriptstyle \pm\, 0.32}$  & $0.41\, {\scriptstyle \pm\, 1.43}$ \\
Quality Switches $\downarrow$ & $0.00\, {\scriptstyle \pm\, 0.00}$  & $0.00\, {\scriptstyle \pm\, 0.00}$  & $0.85\, {\scriptstyle \pm\, 0.04}$  & $0.04\, {\scriptstyle \pm\, 0.04}$  & $0.09\, {\scriptstyle \pm\, 0.07}$  & $0.35\, {\scriptstyle \pm\, 0.15}$  & $0.39\, {\scriptstyle \pm\, 0.14}$ \\
Quality Difference $\downarrow$ & $0.00\, {\scriptstyle \pm\, 0.00}$  & $0.00\, {\scriptstyle \pm\, 0.00}$  & $0.18\, {\scriptstyle \pm\, 0.10}$  & $0.23\, {\scriptstyle \pm\, 0.20}$  & $0.21\, {\scriptstyle \pm\, 0.20}$  & $0.11\, {\scriptstyle \pm\, 0.07}$  & $0.17\, {\scriptstyle \pm\, 0.05}$ \\
Buffer level $\uparrow$ & $7.96\, {\scriptstyle \pm\, 0.41}$  & $1.30\, {\scriptstyle \pm\, 1.28}$  & $3.78\, {\scriptstyle \pm\, 3.17}$  & $6.98\, {\scriptstyle \pm\, 0.77}$  & $6.78\, {\scriptstyle \pm\, 1.31}$  & $4.24\, {\scriptstyle \pm\, 2.90}$  & $6.70\, {\scriptstyle \pm\, 2.05}$ \\
Total Playback Time $\uparrow$ & $100.00\, {\scriptstyle \pm\, 0.12}$  & $56.43\, {\scriptstyle \pm\, 34.21}$  & $85.59\, {\scriptstyle \pm\, 22.96}$  & $99.99\, {\scriptstyle \pm\, 0.29}$  & $99.99\, {\scriptstyle \pm\, 0.56}$  & $99.62\, {\scriptstyle \pm\, 2.32}$  & $98.66\, {\scriptstyle \pm\, 8.31}$ \\
\bottomrule
    \end{tabular}
    }
    \label{tab:results}
\end{table}
\begin{figure}[!t]
    \centering
    \begin{tikzpicture}
    \newcommand{\figheight}{3.55cm}
	\node[inner sep=0pt] (a) {\includegraphics[height=\figheight,trim={0.2cm 0 0cm 0},clip]{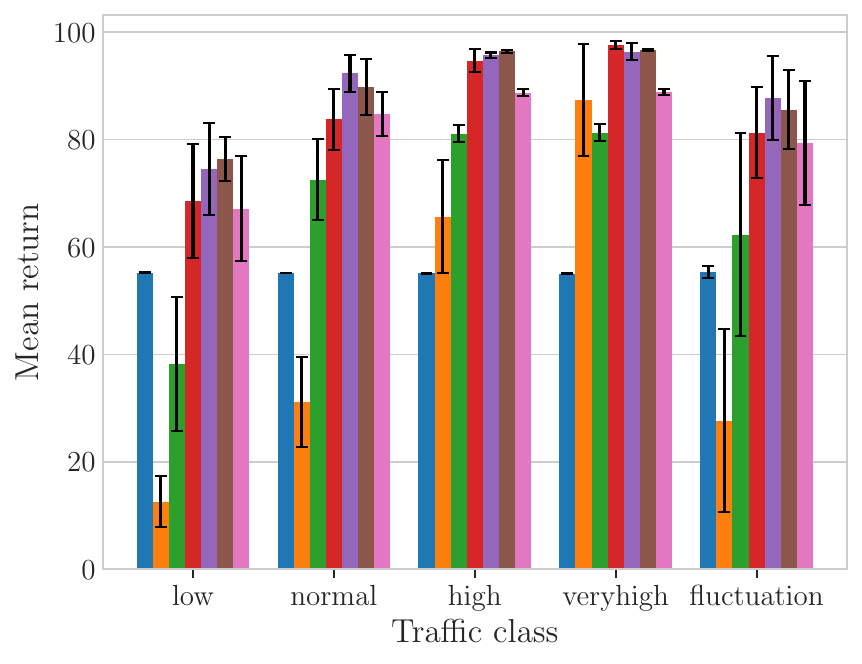}};
	\node[above, inner sep=0pt] at ([shift={(0.125cm,0.0cm)}] a.north) {a) Mean return};
	\node[inner sep=0pt] (b) [right= 0.1cm of a] {\includegraphics[height=\figheight,trim={0 0 0cm 0},clip]{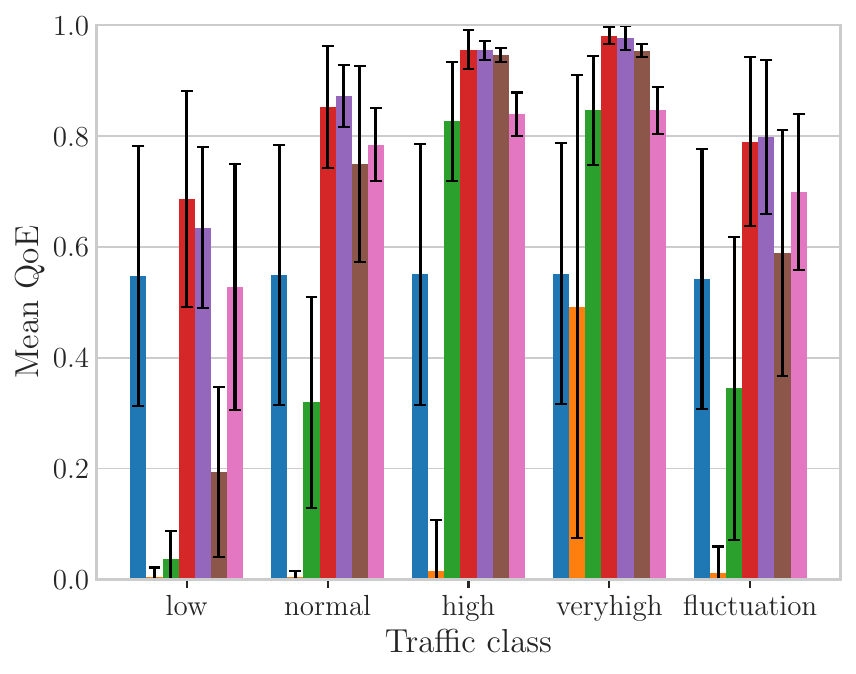}};
	\node[above, inner sep=0pt] at ([shift={(0.125cm,0.0cm)}]b.north) {b) Mean QoE};
	\node[inner sep=0pt] (c) [right=0.1cm of b] {\includegraphics[height=\figheight,trim={0 0 0cm 0},clip]{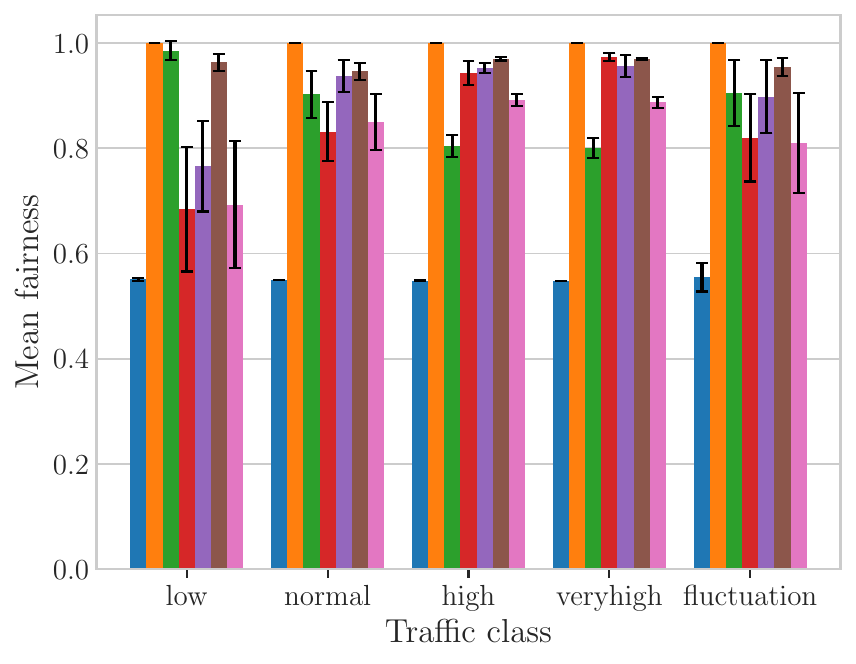}};
	\node[above, inner sep=0pt] at ([shift={(0.125cm,0.0cm)}]c.north) {c) Mean fairness};
	\node[inner sep=0pt] (d) [below= 0.1cm of b] {\includegraphics[height=0.4cm,trim={0cm 8.5cm 0 0.4cm},clip]{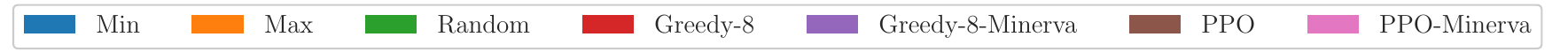}};
    \end{tikzpicture}
    \caption{Mean a) return, b) \ac{qoe} and c) fairness on the test traces when streaming with the four heterogeneous clients Phone, HDTV, 4KTV, and PCV. The results of all agents are aggregated per traffic class, the standard deviation across all traces is indicated by the error bars. The \ac{ppo} agent is outperformed by the greedy baselines and shows a comparatively low \ac{qoe} for the low, normal and flunctuation classes in subplot b).}
    \label{fig:results-overview}
\end{figure}
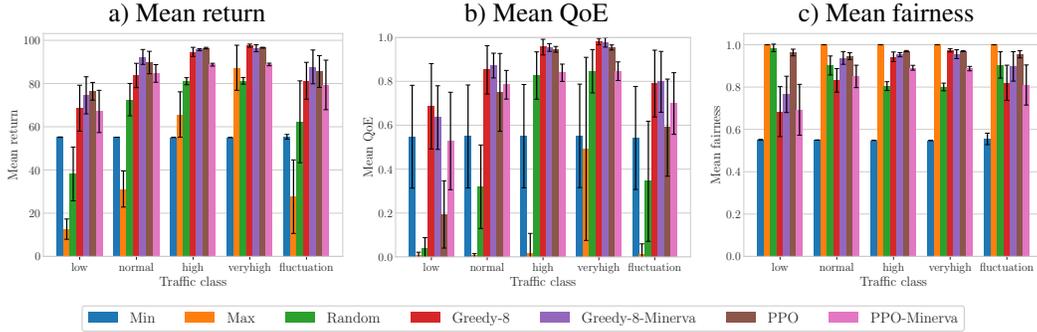

\subsection{Evaluation on the Test Traces}
\label{sec:eval-overview}

The performance of all approaches on the test traces is summarized in Tab.~\ref{tab:results} and Fig.~\ref{fig:results-overview}.
The parameter of the Greedy-$k$ agent was set to $k=8$ as a trade-off between adaptivity and stability. More details regarding this choice are in the appendix, see Sec.~\ref{appendix:greedy-param}.
While the Min agent allows streaming with nearly no interruptions, the corresponding \ac{qoe} varies across the different clients, resulting in a high \ac{qoe} standard deviation in Fig.~\ref{fig:results-overview} b) and a low fairness in Fig.~\ref{fig:results-overview} c).
Conversely, the Max agent is very close to maximum fairness.
However, this is accompanied by a very low \ac{qoe} due to excessive rebuffering.
In the veryhigh trace class, the \ac{qoe} of the Max agent shows a high standard deviation.
This is because some traces from the veryhigh class allow all clients to stream at the highest quality, while others have insufficient bandwidth.
The random agent outperforms the Min and Max agents in terms of reward in four out of five classes.
The \greedy agent outperforms the random agent in terms of reward on all traffic classes.
Although it shows a lower fairness for the low traffic class, this is accompanied by a much higher mean \ac{qoe}.
The \gminerva agent slightly improves upon the \greedy agent, in particular with respect to fairness.

The shown results for the learning-based agents \ac{ppo} and \ac{ppo}-Minerva are the best out of three training runs. 
Details regarding the used hyperparameters and stability across the training runs are provided in appendix Sec.~\ref{appendix:ppo-details}.
Fig.~\ref{fig:results-overview} a) shows that the return of both learning approaches is better than Random, but they fall short of the Greedy approaches.
Both learning approaches have a much lower \ac{qoe}, but the fairness of \ac{ppo} is higher than the one of the Greedy approaches.
In turn, \ac{ppo} performs particularly poor in terms of mean \ac{qoe} in the low traffic class.
Tab.~\ref{tab:results} shows that both learning approaches have much higher rebuffering times than the Greedy approaches, which would be highly undesireable in practice.
This suggests that one of the reasons for the low \ac{qoe} of the learning approaches is that they occasionally accept low quality caused by rebuffering, as this leads to high fairness.

\subsection{Validation Results during Training}
\label{sec:eval-training}

\begin{figure}[!t]
    \centering
    \begin{tikzpicture}
	\newcommand{\figheight}{4.2cm}
	\node[inner sep=0pt] (a) {\includegraphics[height=\figheight,trim={0 0 0cm 0},clip]{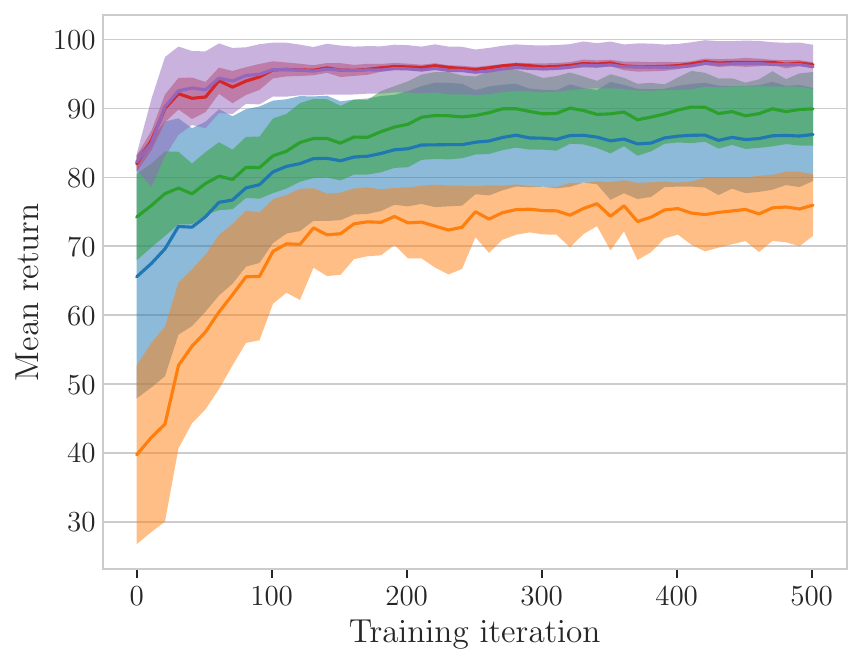}};
	\node[above, inner sep=0pt] at ([shift={(0.125cm,0.0cm)}] a.north) {a) Training return};
	\node[inner sep=0pt] (b) [right= 1cm of a] {\includegraphics[height=\figheight,trim={0 0 0cm 0},clip]{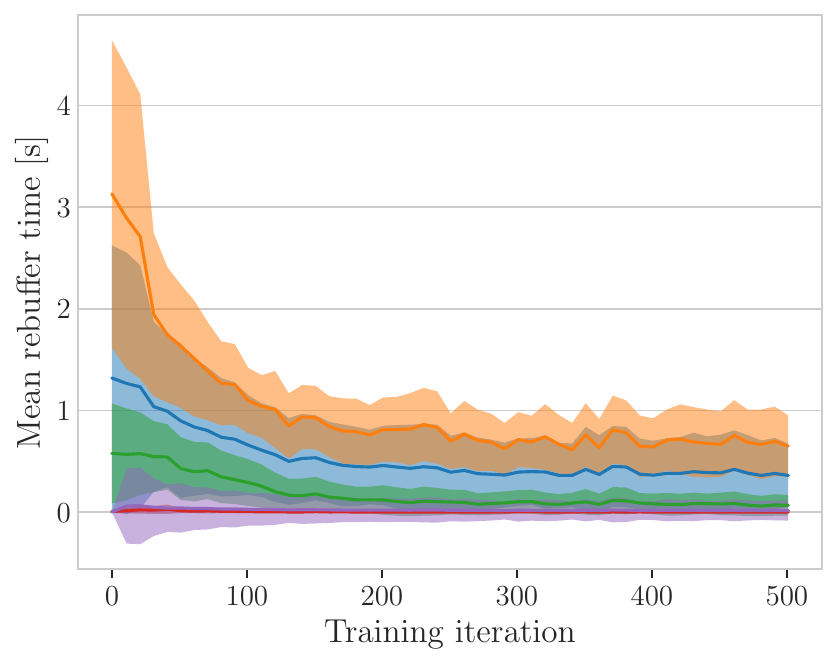}};
	\node[above, inner sep=0pt] at ([shift={(0.125cm,0.0cm)}]b.north) {b) Training rebuffer time};
	\node[inner sep=0pt] (c) at ($(a)!0.5!(b) - (0, 2.4cm)$) {\includegraphics[height=0.65cm,trim={0.2cm 8.5cm 0.4cm 0cm},clip]{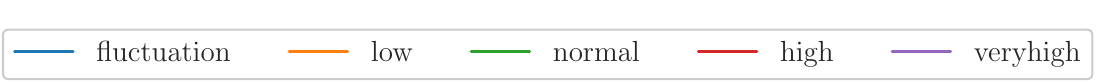}};
    \end{tikzpicture}
    \caption{Mean return a) and rebuffer time b) per traffic class during training of the \ac{ppo} agent, evaluated on the validation traces. The shaded areas show the standard deviation across all validation traces of the respective class. While the returns for the low, normal, high and very high classes are relatively stable, the fluctuating (blue) class shows a high standard deviation. The agent is unable to avoid rebuffering on the low and fluctuation traffic classes.}
    \label{fig:val-training}
\end{figure}

Fig.~\ref{fig:val-training} shows the return and mean rebuffer time of the \ac{ppo} agent on the validation traces during training.
All traffic classes show a noticeable increase in reward over the first 100 training iterations, the results afterwards are comparatively stable.
The standard deviation for the low, normal, and fluctuation traffic classes is considerably higher than for the high and veryhigh traffic classes. This is consistent with the test results in Fig.~\ref{fig:results-overview} a).
The rebuffering time on the high and veryhigh traces is close to zero and the standard deviation on the veryhigh traces decreases during training.
While the rebuffering time for the remaining traffic classes low, normal, and fluctuation decreases considerably during training, the \ac{ppo} agent is unable to reduce it to zero.

\subsection{Effect of Heterogeneous Clients}
\label{sec:eval-clients}

\begin{figure}[!t]
    \centering
    \begin{tikzpicture}
	\newcommand{\figheight}{4cm}
	\node[inner sep=0pt] (a) {\includegraphics[height=\figheight,trim={0 0 0cm 0},clip]{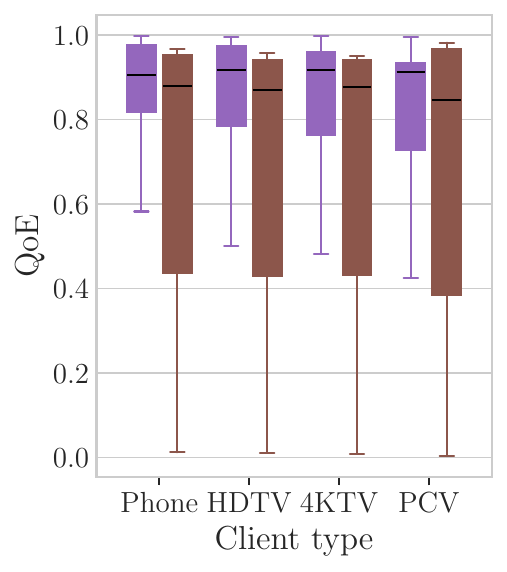}};
	\node[above, inner sep=0pt] at ([shift={(0.125cm,0.0cm)}] a.north) {a) Client \acs{qoe}};
	\node[inner sep=0pt] (b) [right= 1cm of a] {\includegraphics[height=\figheight,trim={0 0 0cm 0},clip]{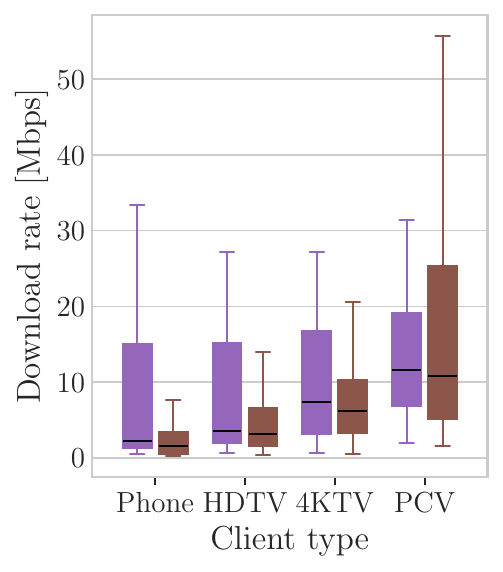}};
	\node[above, inner sep=0pt] at ([shift={(0.125cm,0.0cm)}]b.north) {b) Client download rate};
	\node[inner sep=0pt] (c) at ($(a)!0.5!(b) - (0, 2.2cm)$) {\includegraphics[height=0.6cm,trim={0.2cm 6.9cm 0.3cm 0cm},clip]{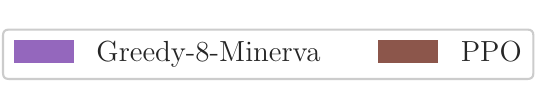}};
    \end{tikzpicture}
    \caption{Mean \ac{qoe} and selected bitrates of \gminerva and \ac{ppo} agents for each client type, aggregated over all traffic classes. The boxes show the range from the first to the third quantile, the median is shown as a black line. The whiskers extend to the farthest point within $1.5$ of the inter-quantile range. Both agents show higher download rates with increasing resource requirements, \ac{ppo} agents show a higher variance in \ac{qoe}.}
    \label{fig:results-device-differences}
\end{figure}

The previous sections have shown that the performance of the agents varies across the different traffic classes.
Additionally, the agents have heterogeneous requirements.
Fig.~\ref{fig:results-device-differences} shows the mean \ac{qoe} and the mean selected bitrate by the \gminerva and \ac{ppo} agents on the test traces.
On the left side, we can see that the median \ac{qoe} of the \ac{ppo} agent is slightly lower than the median \ac{qoe} of the \gminerva agent, while showing a much higher variance.
This is an indication for rebuffering of the agents, which is consistent with the findings from the previous subsections.
On the right side, we can see that the download rate of both approaches increases from left to right with incrementally higher resource requirements.
While the variance of the \gminerva agent is rather consistent across the different clients, the \ac{ppo} agent shows a high variance for the PVC client and a comparably low variance for the other client types.

\subsection{Behavior for an Exemplary Test Trace}
\label{sec:eval-trace}
\begin{figure}[!t]
    \centering
    \begin{tikzpicture}
    \newcommand{\figheight}{4.3cm}
	\node[inner sep=0pt] (a) {\includegraphics[height=\figheight]{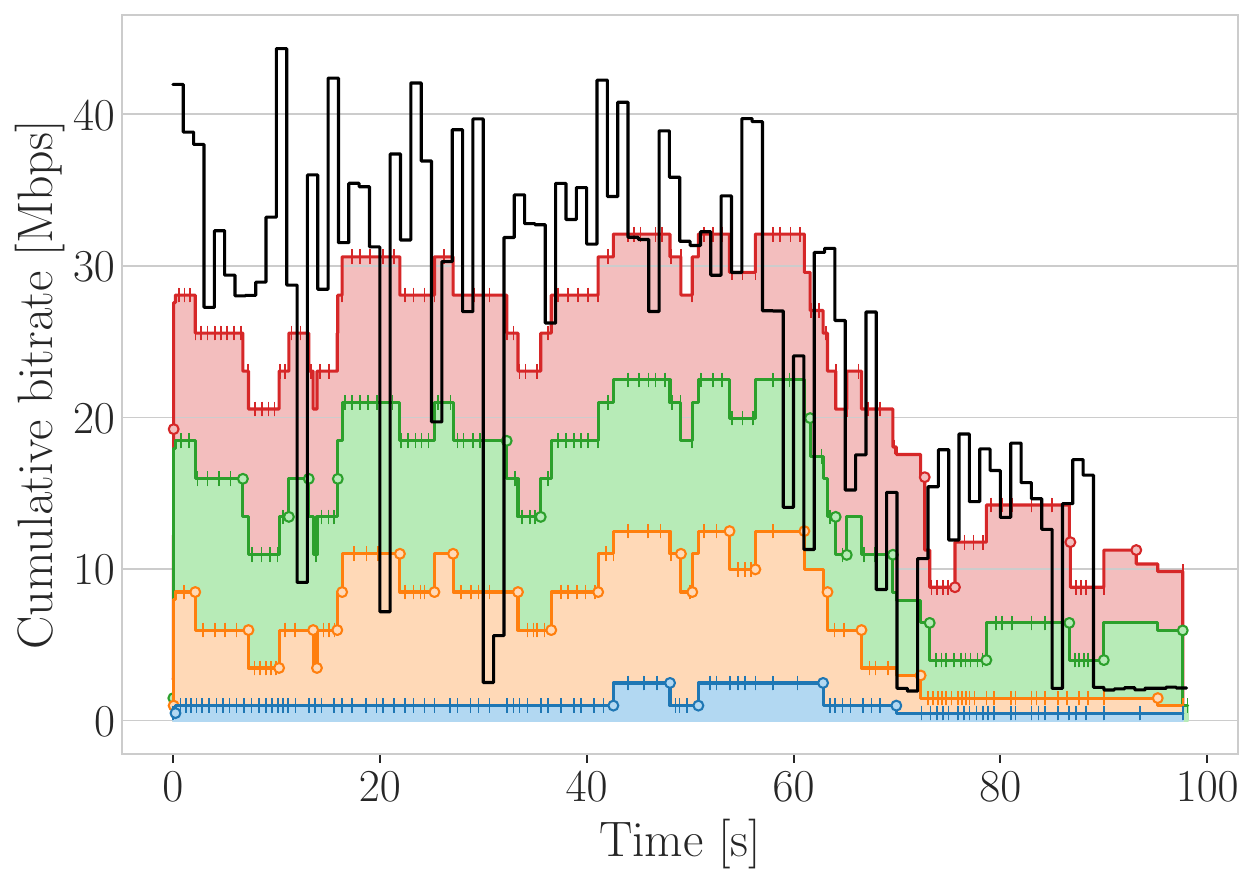}};
	\node[above, inner sep=0pt] at ([shift={(0.125cm,0.0cm)}] a.north) {a) \gminerva};
	\node[inner sep=0pt] (b) [right= 0.5cm of a] {\includegraphics[height=\figheight]{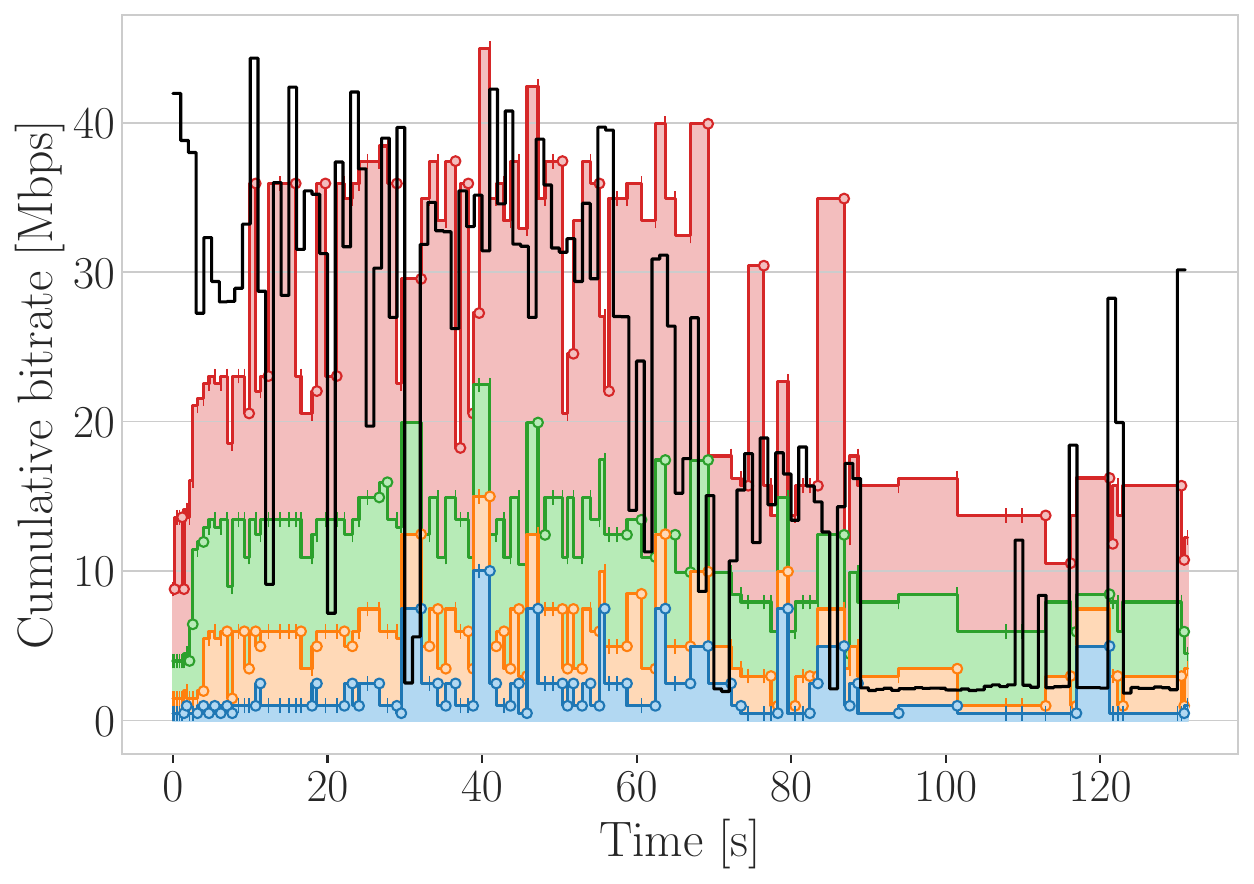}};
	\node[above, inner sep=0pt] at ([shift={(0.125cm,0.0cm)}]b.north) {b) PPO};
	\node[inner sep=0pt] (c) at ($(a)!0.5!(b) - (0, 2.5cm)$) {\includegraphics[height=0.45cm,trim={0cm 11cm 0cm 0.4cm},clip]{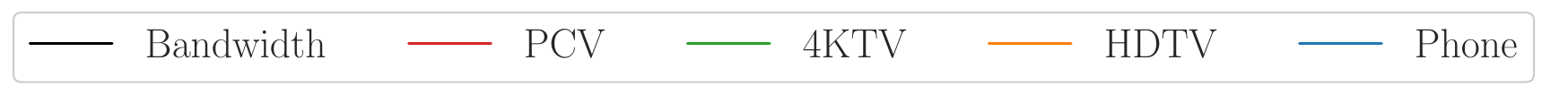}};
    \end{tikzpicture}
    \caption{Behavior of a) \gminerva and b) PPO agents on an exemplary test trace from the fluctuation class. The total bandwidth is shown as a black line. The colored lines and stacked areas indicate the cumulative bitrate of the four clients. Each vertical dash on a line shows an agent step and each dot represents a quality switch. The \gminerva agent adapts well to the declining bandwidth, while the \ac{ppo} agents' high bitrate leads to rebuffering and a longer total download time (see different scale of the time axes).}
    \label{fig:trace-comparison}
\end{figure}

Fig.~\ref{fig:trace-comparison} shows the behavior of \gminerva and \ac{ppo} for an exemplary test trace from the fluctuation class.
The \gminerva agents react swiftly to the declining bandwidth at around $60$ seconds and finish downloading below $100$ seconds with very little rebuffering. The \ac{ppo} agents also reduce their bitrate at around $70$ seconds, but react too greedily to short bandwidth spikes. This leads to a high amount of rebuffering. 
In this particular example, downloading $100$ seconds of multimedia content takes more than $120$ seconds with the \ac{ppo} agents.

\subsection{Effect of Quality-Fairness Coefficient}
\label{sec:ppo-effect-fairness-coeff}
The previous results for the \ac{ppo} agent are for a quality-fairness coefficient $\alpha = 0.25$.
Fig.~\ref{fig:ppo-fairness-coeff} shows these results in comparison with \ac{ppo} agents trained using $\alpha = 0.5$ and $\alpha = 0.75$.
We can see that none of the selected values for $\alpha$ leads to behavior that clearly dominates the others in terms of \ac{qoe} and fairness.
As expected, a higher $\alpha$ tends to increase the \ac{qoe} at the cost of decreased fairness.
One side-effect of increasing the \ac{qoe} is that $\alpha = 0.5$ and $\alpha = 0.75$ show lower rebuffering times during playback, as rebuffering drastically reduces the \ac{qoe}.
However, they still show rebuffering during playback except for the high and veryhigh traffic classes.
In these classes, even \ac{ppo} with $\alpha = 0.25$ was able to achieve very low rebuffering times.
At the same time, the initial rebuffering times for $\alpha = 0.5$ and $\alpha = 0.75$ increase as the agents presumably select higher initial bitrates to reduce the negative effect of the switching penalty.
This shows that agents find yet another undesirable trade off between two factors in the \ac{qoe} definition in order to maximize their return. 

\section{Discussion}
\label{sec:discussion}

\begin{figure}[!t]
    \centering
    \includegraphics[width=0.5\linewidth]{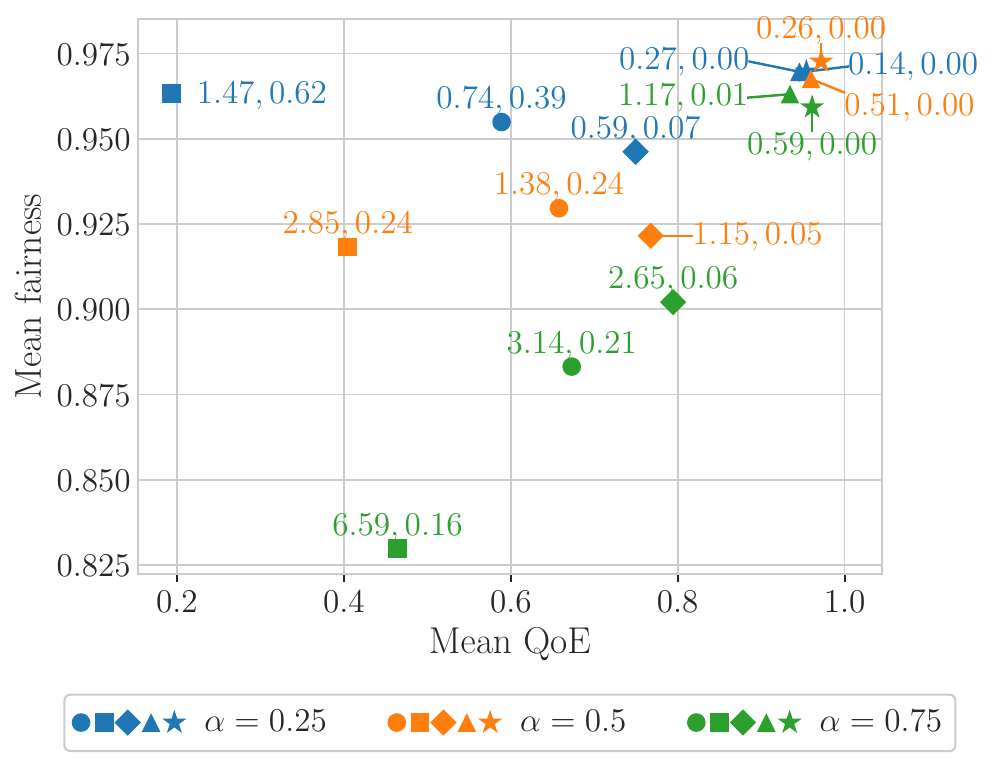}
    \caption{\ac{qoe} and fairness of the \ac{ppo} agent trained with different quality-fairness coefficients $\alpha \in \{0.25, 0.5, 0.75\}$, as indicated by the colors. For each configuration, the best out of three runs in terms of mean return is shown. The markers show separate results for the five traffic classes fluctuation ($\medblackcircle$), low ($\medblacksquare$), normal ($\medblackdiamond$), high ($\medblacktriangleup$) and veryhigh ($\medblackstar$). The text at each marker shows the respective initial stalling time and mean stalling time per segment.}
    \label{fig:ppo-fairness-coeff}
\end{figure}

The results from Sec.~\ref{sec:experiments} demonstrate that a simple greedy heuristic based on the previously measured bandwidths can achieve comparably good results on average.
A combination with the Minerva bandwidth sharing approach from \citet{nathan19minerva} leads to even higher returns due to increased fairness.
Our experiments show that this is a tough baseline for learning approaches.
While the \ac{ppo} agent outperforms the \greedy baseline without Minerva, it has high rebuffering times in three out of five traffic classes and would be unusable in practice.
In particular, we find that the agent is too greedy with respect to changes in the bitrate.
Combining the \ac{ppo} agent with Minerva leads to worse results.
A potential reason for this is that agents with the proportional bandwidth allocation are synchronous, while Minerva bandwidth allocation introduces asynchronicity. 

With a quality-fairness coefficient of $\alpha = 0.25$, the \ac{ppo} agent prioritizes fairness over \ac{qoe} and fails to reach acceptable \ac{qoe} on many traces due to a high amount of rebuffering.
The reason for this is that a low \ac{qoe} caused by rebuffering leads to high fairness, as illustrated by the Max agent in Fig.~\ref{fig:results-overview}.
While the Max agent has poor returns due to clearly suboptimal behavior, the \ac{ppo} agent learns a trade off between \ac{qoe} and fairness that is effective in terms of return, but undesirable in practice.
Sec.~\ref{sec:ppo-effect-fairness-coeff} suggest that simply changing the value of $\alpha$ does not suffice, as even $\alpha=0.75$ shows rebuffering with simultaneously decreased fairness.
The high initial rebuffering times for $\alpha=0.75$ in Fig.~\ref{fig:ppo-fairness-coeff} suggests that increasing the initial rebuffering penality coefficient $\lambda_\text{init}$ could help.
However, when assuming that this \ac{qoe} function would correctly represent the subjective human perception, manually tweaking its parameters to get better agents is not desirable in practice.

To improve upon the naive \ac{ppo} agent considered in this work, we think that two main aspects have to be addressed by future \ac{rl} agents:
\begin{enumerate}
    \item One major limitation of our \ac{ppo} agent is its fixation on a specific value of $\alpha$. Instead of directly weighting the objectives as in Eq.~\ref{eq:combined-objective-reward}, employing multi-objective \ac{rl} with separate estimators for each objective~\cite{hayes2022practical} would be promising. One could also treat the individual components of the \ac{qoe} (see Sec.~\ref{sec:metrics-qoe}) as separate objectives. This would allow to better investigate the policy space and select the best policy, similar to Sec.~\ref{sec:quality-fairness-coefficient-choice} for the time-independent case. It would even allow to change the weights of the objectives at runtime, e.g.~based on preferences by content providers or users. This would lead to more generally applicable agents.
    \item The \ac{ppo} agent is decentralized during execution and only works in one specific setup. An important next step is to introduce joining and leaving clients at any time. However, to achieve fairness, clients have to be aware of other clients in the network. This requires some form of coordination between the clients, which could be achieved with a centralized coordinator~\cite{georgopoulos13openFlowQoEFair, subhan24jointQoEdrl, yuan24qoeFairHeterogeneousMobile} or via communication between clients. A promising direction in that regard is learned communication~\cite{zhu2022survey}. 
    However, research in this direction typically assumes synchronous agents and has to be extended to the asynchronous case.
\end{enumerate}

Additionally, we think that the following extensions of the environment are worth investigating:
\begin{enumerate}
    \item We use predetermined bandwidth sharing modes (see Sec.~\ref{sec:env-bandwidth}), which do not necessarily lead to optimal behavior. While \gminerva benefits from the Minerva bandwidth sharing approach, the \ac{ppo} agent shows poor performance. It would be interesting to explore further bandwidth sharing schemes, particularly when they are designed with \ac{qoe} fairness in mind~\cite{su24qoefairnessHeterogeneousCongestionControl}. Letting each agent select their own weight for bandwidth sharing would also be an option, i.e.~expanding the action space by a continuous action for the weight.
    Alternatively, allowing agents to pause downloads would even enable them to control their bandwidth share under TCP-fair conditions. 
    \item In our environment, if agents select a high bitrate and the bandwidth drops significantly, they are forced to finish downloading the segment. This leads to very high download times and potentially rebuffering. Letting agents cancel the download of a segment at any time would allow them to better react to bandwidth changes and correct previous decisions.
    \item Expanding upon the previous point, giving agents the ability to overwrite segments in their buffer in a non-sequential manner could further improve performance. Intuitively, this would allow agents to first fill their buffers with low and comparatively safe bitrates to ensure playback without stalling, and then selectively increase the quality if this is possible.
\end{enumerate}

\section{Related Work}
\label{sec:related-work}

The application of \ac{rl} in communication networks covers various domains and is expected to play an essential role in the future internet~\cite{luong19DRLApplications, li2022MARLinFutureInternet}.
This work focuses on adaptive bitrate control for multimedia streaming, which has been extensively studied with traditional heuristics and deep \ac{rl} approaches~\cite{bentaleb19abrSurvey}.
\citet{mao17Pensieve} and \citet{gadaleta17ddash} consider a single client that interacts with a streaming environment.
Leveraging the popular \ac{rl} algorithms \acl{a3c}~\cite{mniha2016A3C} and \acl{dqn}~\cite{mnih2015DQN}, they show that learning-based \ac{abr} approaches can outperform previous \ac{abr} algorithms under various network conditions.

When considering multiple clients and shared resources in a streaming system, the objective commonly contains a metric that is shared across all clients, e.g.\ fairness or joint throughput~\cite{georgopoulos13openFlowQoEFair}.
While carefully designed \ac{abr} heuristics address the trade-off between the \ac{qoe} of individual clients and the fairness across all clients~\cite{nathan19minerva, stefano15qoeRateAdaptation, seufert19tcpal}, recent related work started to explore whether learning-based approaches can yield further improvements in the multi-agent setting~\cite{subhan24jointQoEdrl}.
For example, \citet{saaid20dashfair} propose a \ac{rl}-based \ac{abr} approach that controls streams to multiple clients via a centralized server in order to improve the clients' individual \ac{qoe} and the fairness across clients.
Instead of learning the \ac{abr} policy directly, they let agents synchronously select the set of bitrates that a given \ac{abr} algorithm can choose from and show that their method outperforms previous approaches.
\citet{xueqiang24quic} consider streaming from a different perspective, where agents represent multiple QUIC paths of a multimedia streaming application with a single client.
The agents in this environment are heterogeneous, as the paths may utilize different transmission mechanisms.
Considering a combination of individual and shared objectives across agents, they show that their \ac{marl} approach can outperform previous state-of-the-art scheduling algorithms.


As learning-based approaches are predominately trained and evaluated in simulations, the question arises if these results are representative.
\citet{yan20videoInSitu} argue that in a real-world setting, it is difficult for learning-based \ac{abr} methods to outperform even simple heuristics.
A potential reason is that the experiment setup of current learning approaches does not correctly capture the heavy-tailed trace distributions of the real internet. 
\citet{huang22Zwei} argue that defining the correct weights for individual objectives in learning-based \ac{abr} approaches is challenging, as a linear combination with fixed weights can hardly represent the requirements of all types of traffic.

For the future of \ac{rl} methods for \ac{abr}, we think that it is essential for agents trained in simulations to analyze their behavior with respect to different trace distributions and different weights for individual objectives.
We also find that existing multi-agent approaches for fair streaming come with assumptions that limit their real-world applicability, in particular centralized control, homogeneous agents and synchronous steps.
With our work, we propose a novel multi-agent environment that addresses this research gap and takes a step towards lifting these assumptions. 

\section{Conclusion}
\label{sec:conclusion}

With this paper, we model the problem of fair multimedia streaming and propose a novel multi-agent environment that encompasses the challenges of partial observability, multiple objectives, agent heterogeneity and asynchronicity.
We analyze the optimal solutions of a time-independent version of the problem and show that the problem is particularly challenging for lower bandwidths.
This is in line with the empirical results from our experiments.
We categorize the considered bandwidth traces into five classes and show that the agents' behavior can vary drastically between classes.
While the combination of Minerva~\cite{nathan19minerva} and a greedy heuristic performs well across all traffic classes, we show that a naive \ac{ppo} agent fails to learn behavior that would be acceptable in practice.
In particular, the agent can only avoid rebuffering for traces with comparatively high and stable bandwidth.
We argue that fine-tuning the weights of the objective to achieve subjectively better behavior is counterproductive, as this will likely depend on the given trace distribution.
Instead, we suggest that future learning-based approaches for fair streaming should apply and extend methods from multi-objective \ac{rl} and analyze the influence of each objective.

This opens a multitude of directions for future research.
To the best of our knowledge, there is no existing \ac{marl} algorithm that addresses all challenges of this  environment.
A promising direction is therefore to expand and combine existing algorithms.
In particular, multi-objective \ac{rl} approaches would allow to better analyze the space of learned policies.
To increase the realism, one could consider clients that can start and stop streaming sessions during the episode.
As this will require coordination across clients, expanding \ac{marl} approaches with learned communication to asynchronous environments is also a promising research direction.
Further extensions of the environment include changes to the agent's action space to allow agents to correct and modify their decisions on the fly, and the investigation of different bandwidth sharing modes.

\section*{Acknowledgement}
This work has been funded by the German Research Foundation (DFG) in the Collaborative Research Center (CRC) 1053 MAKI. 

\bibliographystyle{ACM-Reference-Format}
\bibliography{bibliography}

\appendix
\newpage
\section{Appendix}

\subsection{Perceptual Quality}
\label{appendix:perceptual-quality}
The perceptual qualities from Fig.~\ref{fig:client-quality-overview} for the 4K, HD and Phone clients are computed as follows:

\begin{enumerate}
    \item We downloaded the Big Buck Bunny movie as 4K PNG images from \url{http://bbb3d.renderfarming.net/explore.html}.
    \item We encoded the movie with x264 in 16:9 format with vertical resolutions 2160p, 1440p, 1080p, and 720p and target bitrates of $0.5, 1, 2.5, 5, 7.5, 10, \text{and }20$ Mbps.
    \item The score of an encoded video is given by the arithmetic average of the VMAF scores for all frames in the clip. For each client, we computed scores for all resolutions below or equal to the respective reference resolution at all bitrates. The HD and Phone clients use the \texttt{vmaf\_v0.6.1.json} model with 1080p at $20$~Mbps as the reference, and the 4K client uses the \texttt{vmaf\_4k\_v0.6.1.json} model with 2160p at $20$~Mbps as the reference. 
    \item For each client type and target bitrate, we selected the resolution that leads to the highest \ac{vmaf} score.
    \item For each client, this results in a \ac{vmaf} score vector $\vec{v} \in [20, 100]^7$, containing a score for each bitrate. We normalized the scores to range $[0, 1]^7$ with $\frac{\vec{v} - 20}{\max(\vec{v}) - 20}$.
\end{enumerate}

For the \ac{pcv} client, we select $7$ quality settings from a subjective study on the quality of point cloud sequences~\cite{weil2023PointCloudQoEModelling} in the near distance setting.
As the \ac{qoe} values $\vec{p}$ are given in an \acl{acr} scale from $1$ to $5$, we normalize them analogously to the \ac{vmaf}s with $\frac{\vec{p} - 1}{\max(\vec{p}) - 1}$.

\subsection{Effect of the Greedy Parameter}
\label{appendix:greedy-param}

\begin{figure}[!t]
    \centering
    \begin{tikzpicture}
    \newcommand{\figheight}{3.4cm}
	\node[inner sep=0pt] (a) {\includegraphics[height=\figheight,trim={0 0 0cm 0},clip]{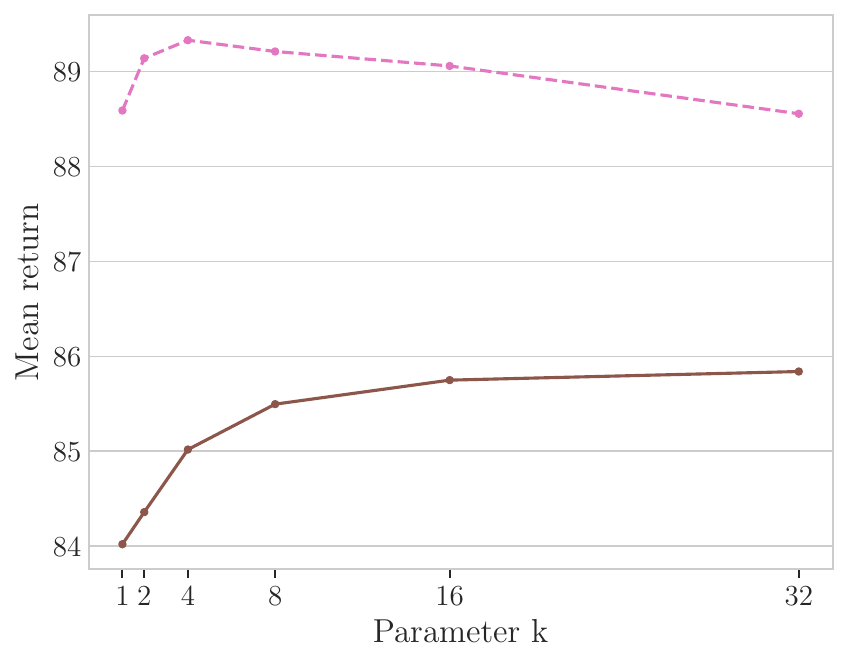}};
	\node[above, inner sep=0pt] at ([shift={(0.125cm,0.0cm)}] a.north) {a) Mean Return};
	\node[inner sep=0pt] (b) [right= 0.25cm of a] {\includegraphics[height=\figheight,trim={0 0 0cm 0},clip]{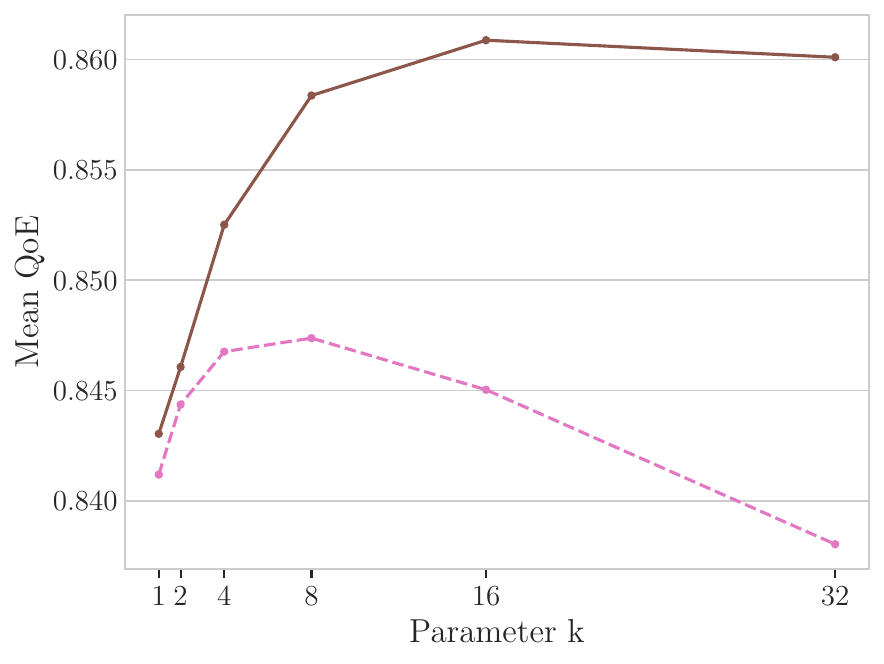}};
	\node[above, inner sep=0pt] at ([shift={(0.125cm,0.0cm)}]b.north) {b) Mean QoE};
	\node[inner sep=0pt] (c) [right=0.25cm of b] {\includegraphics[height=\figheight,trim={0 0 0cm 0},clip]{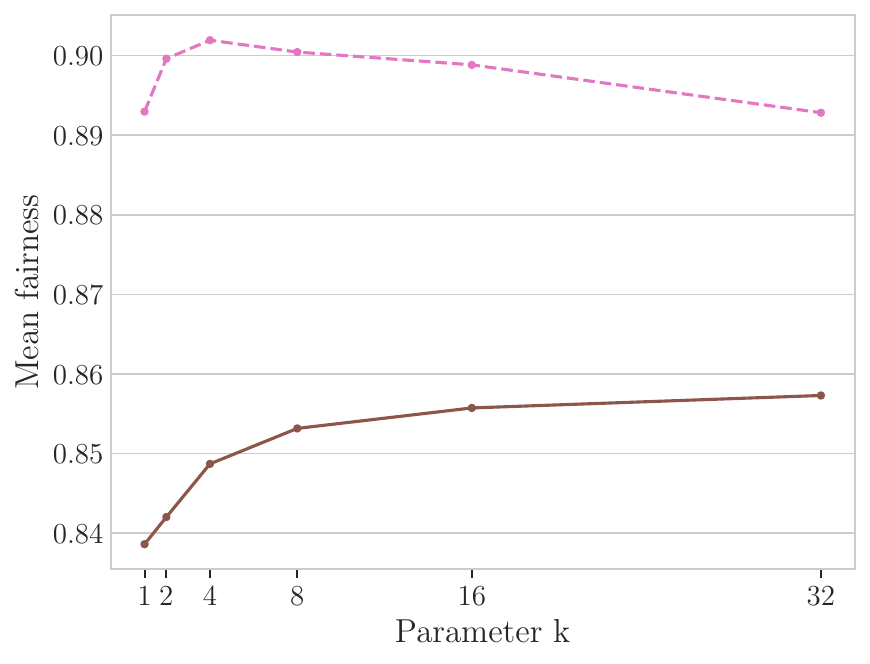}};
	\node[above, inner sep=0pt] at ([shift={(0.125cm,0.0cm)}]c.north) {c) Mean Fairness};
	\node[inner sep=0pt] (xa) [below= 0.1cm of b] {\includegraphics[height=0.5cm,trim={1cm 8.5cm 0 0.4cm},clip]{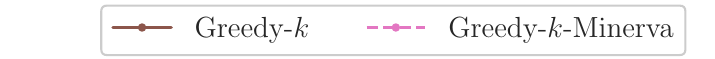}};
	
	\node[inner sep=0pt] (d) [below left= 1.5cm and -1.5cm of b] {\includegraphics[height=\figheight,trim={0 0 0cm 0},clip]{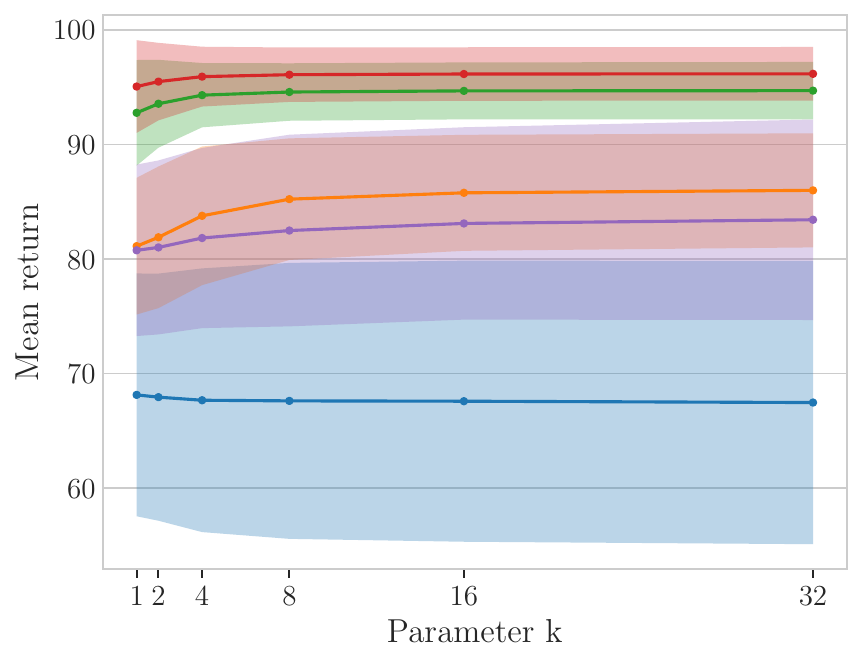}};
	\node[above, inner sep=0pt] at ([shift={(0.125cm,0.0cm)}]d.north) {d) Return of Greedy-$k$ per class};
	\node[inner sep=0pt] (e) [below right= 1.5cm and -1.5cm of b] {\includegraphics[height=\figheight,trim={0 0 0cm 0},clip]{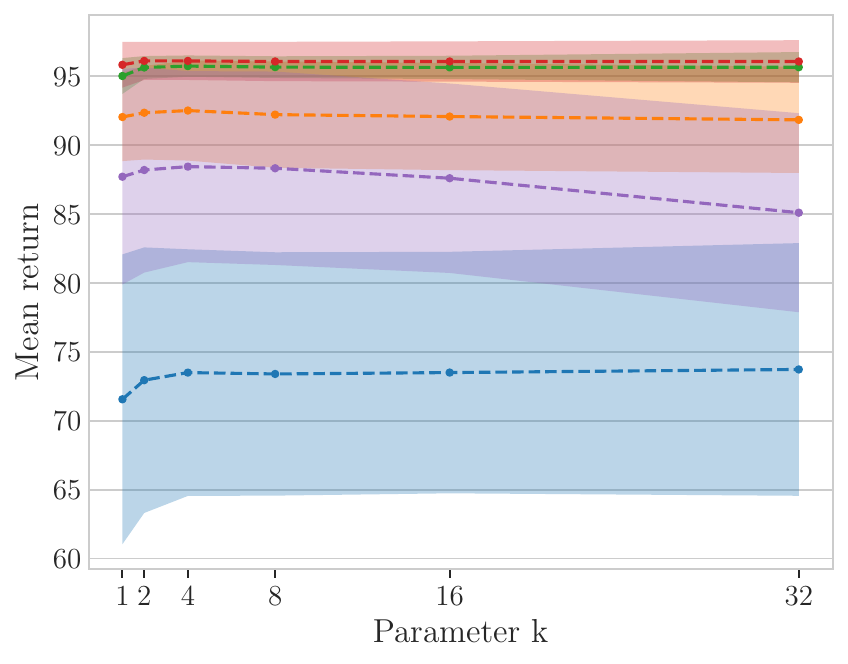}};
	\node[above, inner sep=0pt] at ([shift={(0.125cm,0.0cm)}]e.north) {e) Return of Greedy-$k$-Minerva per class};
	\node[inner sep=0pt] (xb) [below= 5cm of b] {\includegraphics[height=0.5cm,trim={0cm 8.5cm 0.3cm 0.4cm},clip]{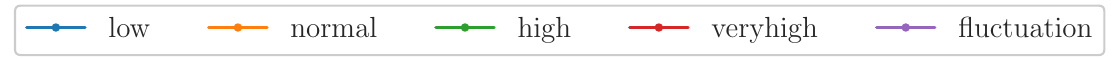}};
    \end{tikzpicture}
    \caption{Mean a) return, b) \ac{qoe} and c) fairness on the validation traces when streaming with the four heterogeneous clients Phone, HDTV, 4KTV, and PCV. Shown are the results for Greedy-$k$ and Greedy-$k$-Minerva for values $k\in \{1,2,4,8,16,32\}$. The subplots d) and e) show the return of each agent for the separate traffic classes. The shaded areas show the standard deviations over the respective traces.}
    \label{fig:results-greedy-k}
\end{figure}
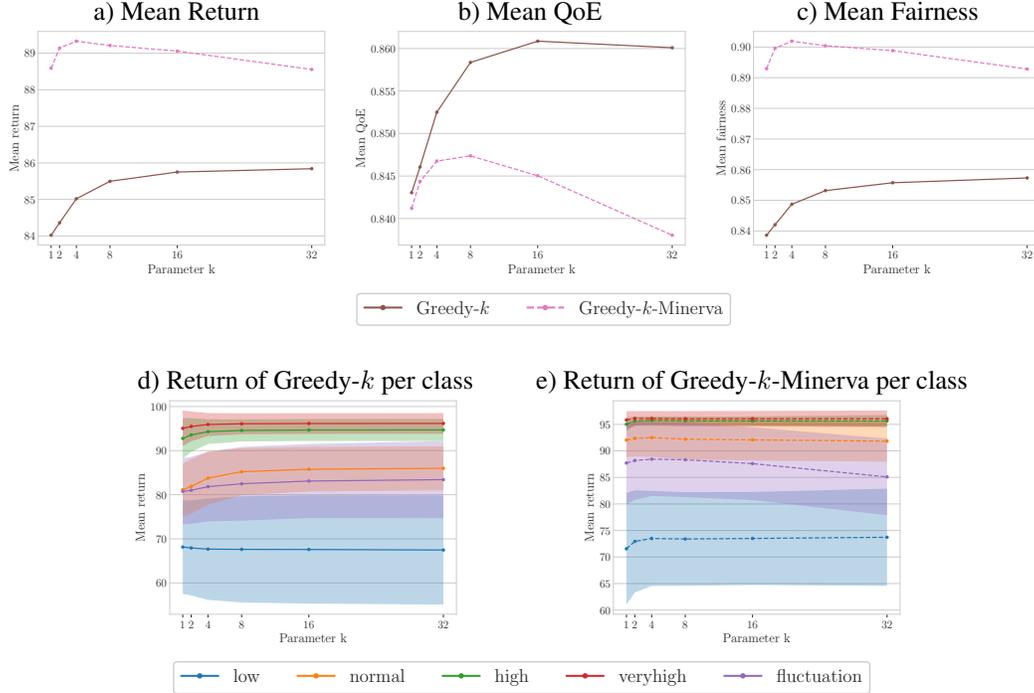

The results of Greedy-$k$ and Greedy-$k$-Minerva for values $k\in \{1,2,4,8,16,32\}$ are shown in Fig.~\ref{fig:results-greedy-k}. 
While the mean return for Greedy-$k$ keeps increasing with a higher value of $k$, the mean return of Greedy-$k$-Minerva decreases after a peak at $k=4$.

A possible explanation for these results is that on average, the traces in the considered data set are rather stable. 
Only the fluctuating class contains traces with rapidly changing bandwidth, rewarding the increasingly conservative behavior of Greedy-$k$ for higher $k$.
However, we can see that the \ac{qoe} starts to decrease from $k=16$ to $k=32$.
Minerva leads to a rather conservative usage of the bandwidth by design, restricting each client to a share of the available bandwidth that would allow them to stream in a fair manner.
Minerva shows more cautious behavior with lower bitrates with increasing $k$, leading to a declining \ac{qoe} after $k=8$.
The return per class in Fig.~\ref{fig:results-greedy-k} d) shows that this decrease is caused by the fluctuation traces, as the returns for the other trace types are rather stable.
Surprisingly, the Greedy-$k$ agent does not suffer from this problem in the flunctuation traces (see subfigure d)).
Its return decreases slightly for the low traces with increasing $k$.

Based on these results, we select the parameter $k=8$ for both approaches for our main evaluation.
While this is not the respective maximum in mean return on the validation trace distribution, it is a compromise between too greedy approaches (low $k$) and too conservative behavior (high $k$).

\subsection{PPO Agent and Ablations}
\label{appendix:ppo-details}

\begin{table}[!t]
    \small
    \centering
    \caption{Training parameters of the \ac{ppo} agent.}
    \begin{tabular}{lc}
\toprule
\textbf{Parameter} & \textbf{Value} \\ \midrule
Learning rate & $1\mathrm{e}{-5}$ \\
Discount factor $\gamma$ & $0.99$ \\
Iterations & $1\:000$ \\
Training batch size & $4\:000$ \\
Minibatch size & $128$ \\
SGD iterations per batch & $30$\\
Evaluation interval (iterations) & $10$\\[3pt]
\hline\\[-1.5ex]
Number of stacked frames & $8$\\
MLP layers & $2$\\
Hidden neurons & $256$\\
Activation function & $\tanh$\\
\bottomrule
    \end{tabular}
    \label{tab:ppo-params}
\end{table}

\begin{figure}[!t]
    \centering
    \includegraphics[width=0.45\linewidth]{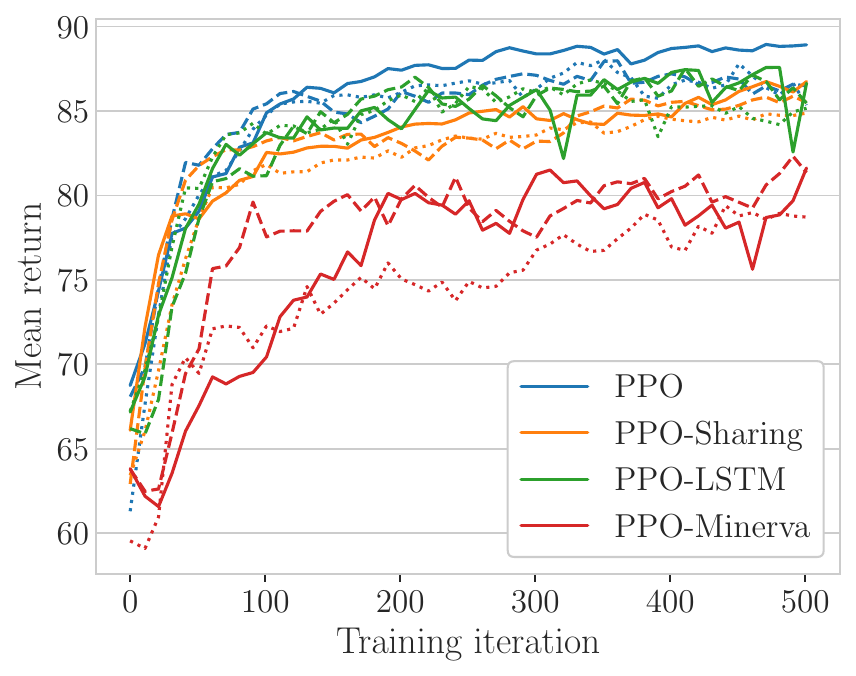}
    \caption{Validation results during training for \ac{ppo} with frame stacking and independent models for each client (PPO), \ac{ppo} with frame stacking and parameter sharing (PPO-Sharing), \ac{ppo} with an LSTM and independent models (PPO-LSTM) and \ac{ppo} with frame stacking, independent models and Minerva bandwidth sharing (PPO-Minerva). For each variant, the plot shows three independent runs with different seeds, with varying linestyle depending on the final return at iteration 500. The respective best run is shown solid~(\rule[.5ex]{1em}{.4pt}), the second run dashed~(\rule[.5ex]{0.47em}{.4pt}\,\rule[.5ex]{.06em}{.0pt}\,\rule[.5ex]{.47em}{.4pt}) and the worst run dotted~($\mathord{\cdot}\mathord{\cdot}\mathord{\cdot}\mathord{\cdot}$).}
    \label{fig:ppo-ablations}
    \vspace{-0.1cm}
\end{figure}

This section provides details regarding the \ac{ppo} agent.

\subsubsection{Parameters}
The parameters are shown in Tab.~\ref{tab:ppo-params}.
In each training iteration, we collect a batch of $4000$ samples by simulating streaming sessions for random traces.
Using this batch, the trainer performs $30$ SGD iterations with minibatch size $128$.
After each $10$ training iterations, we perform a full evaluation of the current model on the validation trace set.

We consider two variations of the agent's architecture, both with two hidden layers and $256$ neurons. The agent from the main paper uses architecture (A) with shared actor and critic networks combined with frame stacking, where the input of the agents consists of the last $8$ observations and actions.
The second architecture (B) is the \rllib default with separate actor and critic networks, and a \ac{lstm} to capture the history of previous observations. In this case, the agents only receive the last observation.

\subsubsection{Stability and Ablations}
Fig.~\ref{fig:ppo-ablations} shows the return of different variations of the \ac{ppo} agent during training.
\ac{ppo}, \ac{ppo}-Sharing and \ac{ppo}-Minerva use architecture~(A), \ac{ppo}-\ac{lstm} uses architecture~(B).
\ac{ppo}, \ac{ppo}-\ac{lstm} and \ac{ppo}-Sharing reach similar returns.
\ac{ppo}-\ac{lstm} is slightly better than \ac{ppo}-Sharing, but is inferior to the \ac{ppo} agent.
The \ac{ppo}-Minerva agents have a noticeably lower mean return and show a high variability across the three training runs.

\begin{figure}[!h]
    \centering
    \begin{tikzpicture}
	\newcommand{\figheight}{4.3cm}
	\node[inner sep=0pt] (a) {\includegraphics[height=\figheight]{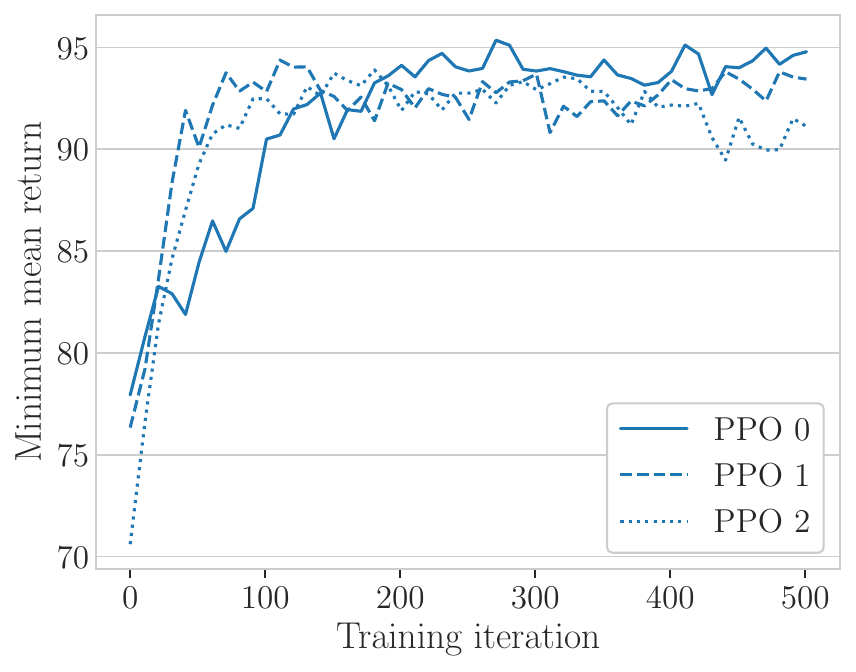}};
	\node[above, inner sep=0pt] at ([shift={(0.125cm,0.0cm)}] a.north) {a) High traffic class};
	\node[inner sep=0pt] (b) [right= 0.5cm of a] {\includegraphics[height=\figheight]{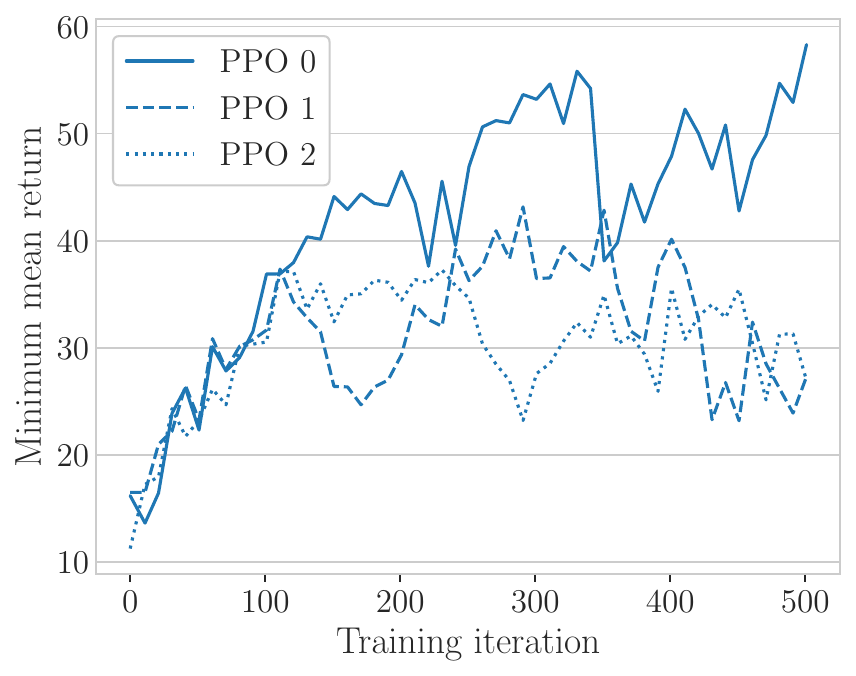}};
	\node[above, inner sep=0pt] at ([shift={(0.125cm,0.0cm)}]b.north) {b) Fluctuation traffic class};
    \end{tikzpicture}
    \caption{Minimum mean return over all validation traces from the a) high and b) fluctuation classes during training for \ac{ppo}. The minimum mean return represents the most challenging trace of the respective class. The linestyle is consistent with Fig.~\ref{fig:ppo-ablations}. The PPO 0 run shows a better minimum mean return in both plots. In b), the minimum mean return of the other two runs PPO 1 and PPO 2 decreases after an intermediate peak.}
    \label{fig:ppo-min}
    \vspace{-0.1cm}
\end{figure}

In Fig.~\ref{fig:ppo-ablations}, the best out of the three \ac{ppo} runs has a considerably higher mean return than the other two runs.
When investigating the individual traffic classes, we notice that all runs show similar results for the high traffic class, but differ for the fluctuation class.
This difference is visible in the worst-case results of the agents, as illustrated in Fig.~\ref{fig:ppo-min}.
In contrast to \ac{ppo} 0, the \ac{ppo} 1 and 2 runs move towards a policy that is suboptimal for the fluctuation traffic traces.

\end{document}